\documentclass[twocolumn,10pt]{article}
\usepackage[top=1.8cm, bottom=1.8cm, left=1.8cm, right=1.8cm]{geometry}
\usepackage{times}  %
\usepackage[hyphens]{url}
\usepackage{graphicx}  %
\usepackage[keeplastbox]{flushend}
\usepackage[hyphens]{url}  %
\usepackage[hang,flushmargin]{footmisc}
\usepackage{xspace}
\usepackage[sort&compress,numbers]{natbib}
\usepackage{multicol, lipsum}
\usepackage[T1]{fontenc}

\usepackage{xcolor}
\usepackage{booktabs}
\usepackage{graphicx}
\usepackage{paralist}
\usepackage[small,bf]{caption}
\usepackage[hang,flushmargin]{footmisc}
\renewcommand{\footnotesize}{\fontsize{8}{9}\selectfont}
\usepackage[raggedright,compact]{titlesec}
\titlespacing*{\section}{0pt}{*4}{4pt}
\titlespacing*{\subsection}{0pt}{*3}{3pt}
\usepackage{xspace}

\makeatletter
\def\url@leostyle{%
  \@ifundefined{selectfont}{\def\UrlFont{}}%
  {\def\UrlFont{}}%
}
\makeatother
\usepackage[hyphenbreaks]{breakurl}

\usepackage[bookmarks=true, bookmarksnumbered=true, colorlinks=true, linkcolor=linkcol, citecolor=citecol, urlcolor=urlcol, hypertexnames=true]{hyperref}

\definecolor{darkgreen}{RGB}{0, 100, 0}
\definecolor{linkcol}{rgb}{0.3,0,0}
\definecolor{citecol}{rgb}{0.3,0,0}
\definecolor{urlcol}{rgb}{0.3,0,0}

\makeatletter
\def\url@leostyle{%
  \@ifundefined{selectfont}{\def\UrlFont{\small}}%
  {\def\UrlFont{}}%
}
\makeatother
\usepackage{amsmath}
\usepackage{amssymb}
\usepackage{graphicx} %
  \setkeys{Gin}{width=.4\textwidth, totalheight=\textheight, keepaspectratio}
  \graphicspath{{graphics/}}
\usepackage[font=footnotesize,labelfont=footnotesize]{subfig}
\usepackage{tikz}
\usepackage{algorithm}
\usepackage{atbegshi}
\usepackage[noend]{algpseudocode}

\usepackage{amsfonts}
\usepackage{color, colortbl}
\usepackage{url}

\newcommand{\Fnaive}{\mathtt{F_{Naive}}}
\newcommand{\Fhist}{\mathtt{F_{Hist}}}
\newcommand{\Fcorr}{\mathtt{F_{Corr}}}
\newcommand{\Fens}{\mathtt{F_{Ens}}}

\newcommand{\tikzcircle}[2][red, fill=red]{\tikz[baseline=-0.5ex]\draw[#1,radius=#2] (0,0) circle ;}%
\definecolor{colorFnaive}{RGB}{31,119,180}
\definecolor{colorFhist}{RGB}{255,127, 14}
\definecolor{colorFcorr}{RGB}{44,160,44}
\definecolor{colorFens}{RGB}{214,39,40}
\newcommand{\descr}[1]{\smallskip\noindent\textbf{#1}}
\newcommand{\codecomm}[1]{}
\newcommand{\codecommb}[1]{}

\usepackage{titling}
\let\OLDthebibliography\thebibliography
\renewcommand\thebibliography[1]{
  \OLDthebibliography{#1}
  \setlength{\parskip}{0pt}
  \setlength{\itemsep}{1pt plus 0.2ex}
}

\begin{document}

\sloppy

\title{\vspace*{-0.75cm}\bf On Utility and Privacy in Synthetic Genomic Data\thanks{Published in the Proceedings of the 29th Network and Distributed System Security Symposium (NDSS 2022).}}

\author{Bristena Oprisanu\\\normalsize UCL\\[-0.5ex]\normalsize bristena.oprisanu.10@ucl.ac.uk \and 
 Georgi Ganev\\\normalsize UCL and Hazy\\[-0.5ex]\normalsize georgi.ganev.16@ucl.ac.uk \and 
 Emiliano De Cristofaro\\\normalsize UCL and Alan Turing Institute\\[-0.5ex]\normalsize e.decristofaro@ucl.ac.uk}
\date{}

\maketitle

\begin{abstract}
The availability of genomic data is essential to progress in biomedical research, personalized medicine, etc. %
However, its extreme sensitivity makes it problematic, if not outright impossible, to publish or share it.
As a result, several initiatives have been launched to experiment with {\em synthetic} genomic data, e.g., using generative models to learn the underlying distribution of the real data and generate artificial datasets that preserve its salient characteristics without exposing it.

This paper provides the first evaluation of both utility and privacy protection of six state-of-the-art models for generating synthetic genomic data.
We assess the performance of the synthetic data on several common tasks, such as allele population statistics and linkage disequilibrium.
We then measure privacy through the lens of {\em membership inference attacks}, i.e., inferring whether a record was part of the training data.
Our experiments show that no single approach to generate synthetic genomic data yields both high utility and strong privacy across the board.
Also, the size and nature of the training dataset matter.
Moreover, while some combinations of datasets and models produce synthetic data with distributions close to the real data, there often are target data points that are vulnerable to membership inference.
Looking forward, our techniques can be used by practitioners to assess the risks of deploying synthetic genomic data in the wild and serve as a benchmark for future work.
\end{abstract}

\section{Introduction}

Progress in genome sequencing has paved the way towards preventing, diagnosing, and treating several diseases and conditions.
Much of this progress is dependent on the availability of genomic data.
Consequently, numerous initiatives have been established to support and encourage genomic data sharing.
For example, the International HapMap Project~\cite{HapMap} helped identify common genetic variations and study their involvement in human health and disease, while the 1000 Genomes Project~\cite{1000g15} created a catalog of human variation and genotype data.
Funding agencies, e.g., the National Institutes of Health (NIH), increasingly often make data sharing a requirement to fund grant applications~\cite{nihgenomicsharing}.

Data sharing in genomics is crucial to enable progress in Precision Medicine~\cite{precisionmedicine}.
However, this inherently conflicts with the need to protect individuals' privacy.
Genomic data contains sensitive information related to heritage, predisposition to diseases, phenotype traits, etc., making it hard to anonymize~\cite{GymrekMGHE13}.
Hiding ``sensitive'' portions of the genome is not effective, as they can still be inferred via high-order correlation models~\cite{SamaniHAEFH15}.\footnote{For a thorough review of privacy threats in genomics, please see~\cite{ayday2015whole,wang2017genome,naveed2015privacy}.} %
As a result, genomics researchers have begun to investigate the possibility of releasing {\em synthetic} datasets rather than real/anonymized data~\cite{synrevolution}.
This follows a general trend in healthcare; for instance, the National Health Service (NHS) in England has recently concluded a project focused on releasing synthetic Emergency Room (``A\&E'') records~\cite{NhsData20}.
Basically, the idea is to use generative models to learn to generate samples with the same characteristics---more precisely, with very close distributions---as the real data.
In other words, rather than releasing data of actual individuals, entities share artificially generated data so that the statistical properties of the original data are preserved, but minimizing the risk of malicious inference of sensitive information~\cite{syntheticperks}.

\descr{Generative Models and Genomics.} In particular, previous work has experimented with both statistical and generative models.
Samani et al.~\cite{SamaniHAEFH15} propose an inference model based on the recombination rate, which can also be used to generate new synthetic genomic samples.
Yelmen et al.~\cite{YelmenDOMMFP19} use Generative Adversarial Models (GANs) and Restricted Boltzmann Machines (RMBs) to mimic the distribution of real genomes and capture population structures. %
Finally, Killoran et al.~\cite{KilloranLDDF17} %
use ad-hoc training techniques for GANs and architectures for computer vision tasks.

\descr{Technical Roadmap.} Prior work on synthetic genomic data (Section~\ref{sec:models}) has not evaluated the privacy guarantees they provide and only evaluated utility from  the shallow point of view of the statistical fidelity provided by the generative model.
To address this gap, we introduce a novel evaluation methodology and perform a series of experiments geared to quantify both the utility and the privacy of six state-of-the-art models used to generate human genomic synthetic data.

Our analysis unfolds along two main axes:
\begin{enumerate}
\item {\em Utility.} We focus on several, very common computational tasks on genomic data, measuring how well generative models preserve summary statistics (e.g., allele frequencies, population statistics) or linkage disequilibrium (see Section~\ref{sec:utility}).
We also compute the Kologorov-Smirnov(KS) test on the main metrics discussed as a non-parametric test.

\item {\em Privacy.} We mount {\em membership inference attacks}~\cite{homer2008resolving}, having an attacker infer whether a target record was part of the real data used to train the model producing the synthetic dataset, as opposed to releasing the real dataset (see Section~\ref{sec:privacy}).
In the process, we also introduce a novel attack where the adversary only has {\em partial} information for a target individual (Section~\ref{sec:new}).  We choose to study MIA as it is one of the most studied attacks mounted for both genomics data and machine learning, however, it has never been studied in the context of synthetic genomic data.
\end{enumerate}

\descr{Main Findings.} Overall, our evaluation shows that there is no single approach for generating genomic synthetic data that performs well across the board, both in terms of utility and privacy.
Among other things, we find that:
\begin{itemize}
\item A high-order correlation model specifically build for genomic data (Recomb) has the best utility metrics for small datasets but does so at the cost of privacy, even against weaker adversaries who only have partial information available.
\item The RBM model has a better performance with increasing dataset sizes, both in terms of utility and privacy, as targets get stronger privacy protection when synthetic data is generated using a larger training set.
\item  Releasing synthetic datasets does not provide robust protection against membership inference attacks.  We find cases where releasing the synthetic dataset sometimes offers better protection against membership inference attacks.
However, because of the randomness introduced by the generative models, one cannot meaningfully predict a target's susceptibility to privacy attacks without fixing the training set and quantifying the respective privacy loss/gain for all targets in the set.
\end{itemize}

\section{Preliminaries}
In this section, we introduce background information about genomics,  and tools used to generate synthetic data, then, we present the privacy metrics and datasets used in our evaluation.

\subsection{Genomic Primer}

\noindent{\bf Genomes and Genes.} The genome represents the entirety of an organism's hereditary information.
It is encoded in DNA: each DNA strand comprises four chemical units, called nucleotides, represented by the letters A, C, G, and T.
The human genome consists of approximately 3 billion nucleotides packaged into thread-like structures called chromosomes.
The genome includes both the genes and the non-coding sequences of the DNA.
The former determine specific traits, characteristics, or control activity within an organism.
We refer to the group of genes inherited from a single parent as a \textbf{\em haplotype}.
An \textbf{\em allele} is a different variation of a gene; any individual inherits two alleles for each gene, one from each of their parents.
The genotype consists of the alleles that an organism has for a particular characteristic.

\descr{SNPs and SNVs.} About 99.5\% of the genome is shared among all humans; the rest differs due to genetic variations.
Single nucleotide polymorphisms (SNPs) are the most common type of genetic variation.
They occur at a single position in the genome and at least 1\% of the population.
SNPs are usually biallelic and can be encoded by \{0, 1, 2\}, with 0 denoting a combination of two major (i.e., common) alleles, 2 a combination of two minor alleles, and 1 a combination of a major and a minor allele (which is also referred to as a heterozygous SNP).
Single nucleotide variants (SNVs) are single nucleotide positions in the genomic DNA at which different sequence alternative exists~\cite{variant}.

\descr{Recombination Rate (RR).} Recombination is the process of determining the frequency with which characteristics are inherited together.
The RR is the probability that a transmitted haplotype constitutes a new combination of alleles different from that of either parental haplotype~\cite{ClarkeC05}.

\descr{Genome-Wide Association Studies (GWAS).} GWAS are hypothesis-free methods for identifying associations between genetic regions and traits.
A typical GWAS looks for common variants in several individuals, both with and without a trait, using genome-wide SNP arrays \cite{gwas, gwas2}.

\subsection{Membership Inference Attacks}\label{sec:mia}

\noindent{\bf Genomics.} A well-understood privacy threat in genomics is determining whether the data of a target individual is part of an aggregate genomic dataset or mixture.
This is known as a {\em membership inference attack} (MIA)~\cite{homer2008resolving, WangLWTZ09, ZhouPLCTW11}. %
The ability to infer the presence of an individual's data in a dataset constitutes an inherent privacy leak whenever the dataset has some sensitive attributes.
For instance, if a mixture includes DNA from patients with a specific disease, learning that a person is part of that mixture exposes their health status.

In general, genomic data contains extremely sensitive information about individuals; %
hence, MIAs against genomic data prompt severe privacy threats, including denial of life/health insurance, revealing predisposition to diseases, ancestry, etc.

\descr{Machine Learning.} MIAs have also been studied in the context of machine learning to infer whether or not a target data point was used to train a target model.
This has been done both for discriminative \cite{ShokriSSS17,SZHBFB19,SDSOJ19,ChenYZF19, hui2021practical,LF20,NSH18}
and generative models~\cite{HayesMDD19,HilprechtHB19,ChenYZF19}.
Inferring training set membership might yield serious privacy violations.
For instance, if a model for drug dose prediction is trained using data from patients with a certain disease, or synthetic health images are produced by a generative model trained on patients' images, learning that data of a particular individual was part of the training set leaks information about that person's health.
Overall, MIAs are also used as signals that access to a target model is ``leaky'' and can be a gateway to additional attacks~\cite{HayesMDD19}.
This is the main reason we choose MIA as our leading privacy metric while deferring the quantification of other attacks (e.g., attribute inference or reconstruction attacks) to future work.

\subsection{Privacy Gain (PG) from Synthetic Data Release}\label{subsec:pg}

Our main privacy evaluation metric is denoted as {\bf\em Privacy Gain (PG)}, first proposed in~\cite{StadlerOT20}.
The PG quantifies the privacy advantage obtained by a target $t$, vis-\`a-vis an MIA adversary, when a synthetic dataset is published instead of the real data.

\descr{MIA Training (Algorithm~\ref{alg:miatrain}).} The adversary is trained as follows.
First, we take a reference dataset (which may or may not overlap with the real data used to generate the synthetic dataset) and use it to generate a synthetic dataset (lines 1--6).
This dataset is labeled as 0, i.e., it does not include the target record (line 7).
The target record is added to the dataset (line 8), and synthetic data is generated from the new dataset, which includes the target record; this dataset is labeled as 1 (lines 10-13).
These models are referred to as {\em generative shadow} models.
Finally, the adversary uses the synthetic datasets to train a classifier (line 14), which distinguishes whether or not the target was used to train a generative model.

\descr{PG Estimation (Algorithm~\ref{alg:miagain}).} We estimate the PG for a fixed target record and input dataset as follows. %
First, the algorithm takes the target training set and generates $n_s$ synthetic datasets without the target record (lines 1--4).
Then, the target record is added to the training set, and $n_s$ of synthetic datasets are generated from this dataset (lines 5--9).
Then, the adversary trains their attack model $\mathtt{MIATrain}$ (line 10).
Finally, the PG for a target $t$ is computed as $ PG_t  = \frac{1 - \overline{MIA_t}(S_{test})}{2}$, where $\overline{MIA_t}(S_{test}) = \sum_{S_i\in S_{test}} \frac{\Pr[MIA_t(S_i) = 1]}{2*n_s}$ (lines 11--12).

Put simply, the PG quantifies the difference between the probability that an attacker correctly identifies that the target record belongs to the real dataset %
and that that the attacker correctly identifies that the target record was used in training a generative model that outputs a synthetic dataset.

\begin{figure*}[t]
\centering
\begin{minipage}[t]{1\columnwidth}
 \begin{algorithm}[H]
	\caption{$\mathtt{MIATrain}$~\cite{StadlerOT20}}
	\label{alg:miatrain}
	\begin{algorithmic}[1]
		\small
		\Require A generative model GM(), the target record $t$, a reference dataset $R$ of size $n$, the number of synthetic test sets $n_s$ of size $m$, and the number $k$ of shadow models. %
		\Ensure$MIA_t()$
		\For{$i = 1, \cdots, k$}
			\State$R_i \sim R^n$ \codecomm{sample a reference dataset}
			\State$f_i \sim$ GM(${R_i}$) \codecomm{train a generative model on the reference dataset without the target}
			\For{$j = 1, \cdots, n_s$}
				\State$S_j^m \sim f_i$ \codecomm{get a synthetic dataset from the trained generative model}
				\State $S_{\mathtt{train}} \gets S_{\mathtt{train}} \cup S_j^m$ \codecomm{add to training set}
				\State$l_{\mathtt{train}}  \gets l_{\mathtt{train}} \cup0$ \codecomm{label as 0}
			\EndFor
			\State$R'_i \gets R_i \cup t$ \codecomm{add target}
			\State$f'_i \sim$ GM(${R'_i}$) \codecomm{\codecomm{train a generative model on the reference dataset with the target}}
			\For{$j = 1, \cdots, n_s$}
				\State$S_j^m \sim f'_i$ \codecomm{get a synthetic dataset from the trained generative model}
				\State$S_{\mathtt{train}} \gets S_{\mathtt{train}}\cup S_j^m$ \codecomm{add to training set}
				\State$l_{\mathtt{train}} \gets l_{\mathtt{train}} \cup1$ \codecomm{label as 1}
			\EndFor
		\EndFor
		\State $MIA_t() {\gets} \mathtt{Classifier}(S_{\mathtt{train}}, l_{\mathtt{train}})$ \codecommb{train MIA classifier}
	\end{algorithmic}
\end{algorithm}
\end{minipage}
\hfill
\begin{minipage}[t]{1\columnwidth}
\begin{algorithm}[H]
	\caption{$\mathtt{MIAGain}$~\cite{StadlerOT20}}
	\label{alg:miagain}
	\begin{algorithmic}[1]
		\small
		\Require A generative model GM(), the target record $t$, the target training set $R^t_{out}$ of size $n$, the size $m$ of the synthetic dataset, the number $n_s$ of synthetic test sets, a reference dataset $R^a$, the number $k$ of shadow models.\vspace{0.25cm}
		\Ensure$PG_t$
		\State$f_{out} \sim $ GM(${R^t_{out}}$) \codecomm{\edc{???}}
		\For{$i = 1, \cdots, n_s$}
			\State$S_i \sim f_{out}^m$ \codecomm{\edc{???}}
			\State$S_{test} \gets S_{test} \cup S_i$ \codecomm{\edc{???}}
		\EndFor

		\State$R^t_{in} \gets R^t_{out} \cup t$ \codecomm{add target}
		\State$f_{in} \sim $ GM(${R^t_{in}}$) \codecomm{\edc{???}}
		\For{$i = 1, \cdots, n_s$}
			\State$S_i \sim f_{in}^m$ \codecomm{\edc{???}}
			\State$S_{test} \gets S_{test}\cup S_i$ \codecomm{\edc{???}}
		\EndFor
		\State$MIA_t() \gets \mathtt{MIATrain}($GM()$, t, R^a, n, m, n_s, k)$
		\State $\overline{MIA_t}(S_{test}) = \sum_{S_i\in S_{test}} \Pr[MIA_t(S_i) = 1]/2n_s$
		\State$PG_t \gets (1 - \overline{MIA_t}(S_{test}))/2$
	\end{algorithmic}
\end{algorithm}
\end{minipage}
\end{figure*}

\descr{PG Values.} The PG ranges between 0, when publishing the synthetic dataset leads to the same privacy loss as publishing the real dataset (i.e., $\overline{MIA_t}(S_{test})$ = 1) and 0.5, when publishing the synthetic dataset perfectly protects the target from MIA (i.e., $\overline{MIA_t}(S_{test})$=0).
This means that $PG=0.25$ when the probability of the adversary inferring whether or not a target is part of the training set used to generate the synthetic dataset is the same as random guessing (i.e., $\overline{MIA_t}(S_{test})=0.5$).

\descr{Dimensionality Reduction.} To reduce the effects of high dimensionality, %
the attacker first maps the synthetic data to a lower feature space.
This helps detect the influence of the target record on the training dataset.
We experiment with four different feature sets, as done in~\cite{StadlerOT20}: namely, a {\em naive} feature set, which encodes the number of distinct categories plus the most and least frequent category for each attribute, a {\em histogram}, which computes the frequency counts for each attribute, a {\em correlation}, which encodes pairwise correlations between attributes, and an {\em ensemble} feature set, which combines all the previously mentioned feature sets.

\subsection{Datasets}

Our evaluation uses data from two projects: HapMap~\cite{HapMap} and the 1000 Genome Project~\cite{1000g15}.
More specifically, we use 1,000 SNPs from chromosome 13 from the following datasets:

\begin{enumerate}

\item {\em CEU Population (HapMap).} Samples from 117 Utah residents with Northern and Western European ancestry, released in phase 2 of the HapMap project.

\item {\em CHB Population (HapMap).} Samples from 120 Han Chinese individuals from Beijing, China.

\item {\em 1,000 Genomes.} Samples from 2,504 individuals from 26 different populations released from phase 3 of the 1000 Genomes project.
\end{enumerate}

\section{Synthetic Data Approaches in Genomics}\label{sec:models}

In this section, we provide an overview of the state-of-the-art models for generating synthetic genomic data.
In particular, we discuss the Recombination model presented by Samani et al.~\cite{SamaniHAEFH15}, the RBM and GAN models proposed by Yelmen et al.~\cite{YelmenDOMMFP19}, and the WGAN model from Killoran et al.~\cite{KilloranLDDF17}.
We also introduce and consider two other ``hybrid'' models.

\descr{Recombination Model (Recomb).}
Samani et al.~\cite{SamaniHAEFH15} propose the use of a {\em recombination model} as an inference method for quantifying individuals' genomic privacy.
This statistical model is based on a high-order SNV correlation that relates linkage disequilibrium patterns to the underlying recombination rate.
\cite{SamaniHAEFH15} shows how to use this method to generate synthetic samples and perform Principal Component Analysis (PCA).
The recombination model yields a distribution closer to the real data than models using only linkage disequilibrium and allele frequencies.
The model uses a ``genetic map,'' which includes the recombination rate.
This is provided with the dataset for the HapMap datasets, but not for the 1000 genomes data.
For the latter, we use the scripts from~\cite{genMaps} to generate the genomic map.

\descr{Restricted Boltzmann Machines (RBMs).}
RBMs~\cite{SmolenskyP86} are generative models geared to learn a probability distribution over a set of inputs.
RBMs are shallow, two-layer neural nets: the first is known as the ``visible'' (on input) layer and the second as the hidden layer.
The two layers are connected via a bipartite graph, i.e., every node in the visible layer is connected to every node in the hidden one, but no two nodes in the same group are connected, allowing for more efficient training algorithms.
The learning procedure consists of maximizing the likelihood function over the visible variables of the model.
The RBM models re-create data in an unsupervised manner through many forward and backward passes between the layers, corresponding to sampling from the learned distribution.
The output of the hidden layer passes through an activation function, which becomes the input for the former.

As mentioned, %
Yelmen et al.~\cite{YelmenDOMMFP19} use RBMs to generate synthetic genomic data.
In our evaluation, we follow the same RBM settings as ~\cite{YelmenDOMMFP19}.
More specifically, we use a ReLu activation function, with the visible layer having the same size as the input we considered (1,000 features) and with the number of hidden nodes set to 100.
The learning rate is set to 0.01, the batch size to 32, and we iterate over 2,000 epochs.

\descr{Generative Adversarial Networks (GANs).}
A GAN is an unsupervised deep learning model consisting of two neural networks, a generator and a discriminator, which ``compete'' against each other. %
During training, the generator's goal is to produce synthetic data, and the discriminator evaluates them against real data samples to distinguish the synthetic from the real samples.
The training objective is to learn the data distribution so that the data samples produced by the generator cannot be distinguished from real data by the discriminator.

We use the GAN approach also proposed by Yelmen et al.~\cite{YelmenDOMMFP19}, mirroring their experimental settings.
That is, the generator model consists of an input layer with latent dimension set to 600 and two hidden layers, of sizes 512 and 1,024, respectively.
The discriminator consists of an input layer with a size equal to the number of SNPs evaluated (1,000) and two hidden layers of sizes 512 and 256, respectively, and an output layer of size 1.
The output layer for the generator uses tanh as an activation function, and the output layer for the discriminator uses the sigmoid activation function.
We compile both the generator and discriminator using the Adam optimization and binary cross-entropy as the loss function. %

\descr{Recombination RBM (Rec-RBM).}
To overcome issues caused by low numbers of training samples, we propose a hybrid approach between the Recomb and the RBM models.
We use the former to generate extra samples, which we then use, together with the real data samples, to train the RBM model with the same parameters as before.
We do so to explore whether having more data points to train the model improves the utility of the synthetic data.

\descr{Recombination GAN (Rec-GAN).}
Like Rec-RBM, we use the Recomb model to generate extra training samples for the GAN model, using the same parameters as before.
Again, we want to study whether having a larger dataset available for training improves the synthetic data output's overall utility.

\descr{Wasserstein GAN (WGAN).}
Killoran et al.~\cite{KilloranLDDF17} propose an alternative GAN model by treating DNA sequences as a hybrid between natural language and computer vision data.
The sequences are one-hot encoded, the GAN is based on a WGAN architecture trained with a gradient penalty~\cite{GulrajaniAADC17}, and both the generator and discriminator use convolutional neural networks~\cite{LeCunBDHHHJ89} and a residual architecture \cite{HeZRS15}, which includes skip connections that jump over some layers.
The authors also propose a joint method combining the GAN model with an activation maximization design \cite{SimonyanVZ14, YosinskiCNFL15, MordvintsevOT15} to tune the sequences to have desired properties.
However, we do not include the joint model in our evaluation, as we focus on a range of statistics instead of a single desired property.

In our evaluation, we use the WGAN model with the default parameters from the implementation in~\cite{genDNA}.
The generator consists of an input layer with a dimension of the latent space set to 100, followed by a hidden layer with a size 100 times the length of the sequence (1,000), which is then reshaped to (length of the sequence, 100), followed by 5 resblocks.
Finally, there is a 1-D convolutional layer followed by the output layer, which uses softmax.
The discriminator has a very similar architecture but in a different order -- i.e., it starts with the input layer to which the one-hot sequences are fed, that is followed by the 1-D convolutional layer, then the 5 resblocks, followed by the reshape layer and the output layer of size 1.
We perform 5 discriminator updates for every generator update.
Both the generator and discriminator use Adam optimization, and their learning rates are set to 0.0001, while a gradient penalty adjusts the loss as mentioned.
We use a batch size of 64.
For more details please refer to Appendix~\ref{app:wgan}.

\begin{figure*}[t]
\centering
\subfloat[CEU]{\includegraphics[width=0.95\textwidth]{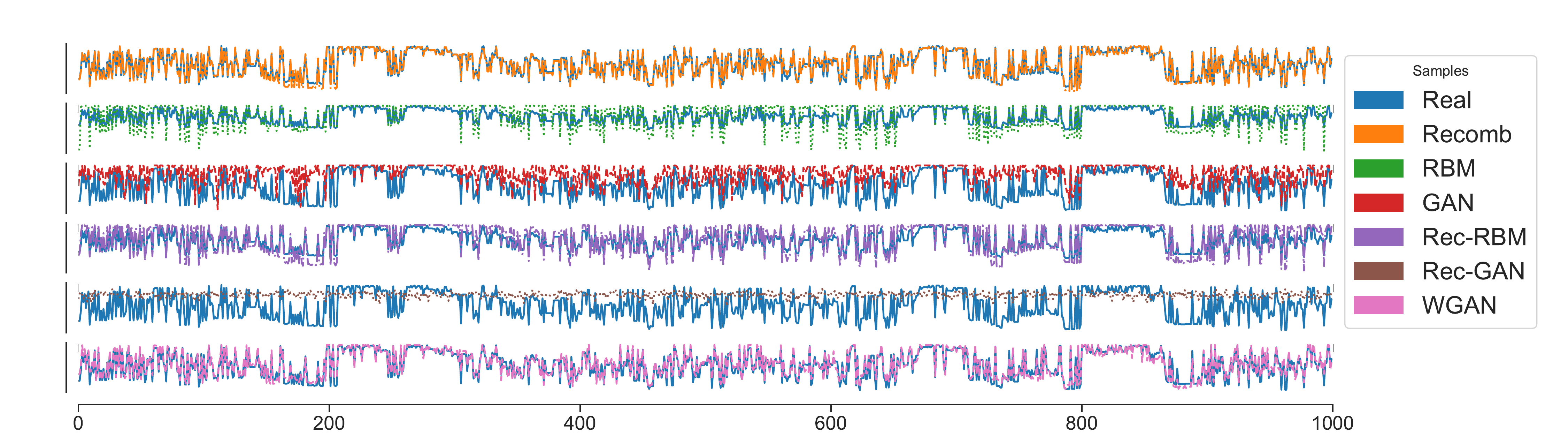}\label{fig:mafceu}}\\[-0.5ex]
\subfloat[CHB]{\includegraphics[width=0.95\textwidth]{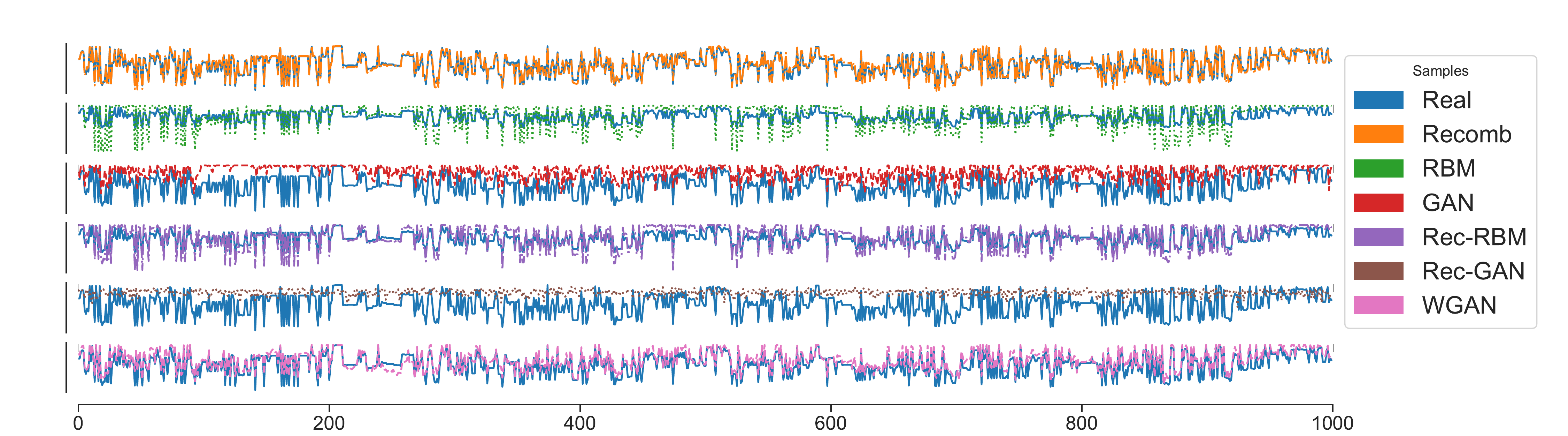}\label{fig:mafchb}}\\[-0.5ex]
\subfloat[1000 Genomes]{\includegraphics[width=0.95\textwidth]{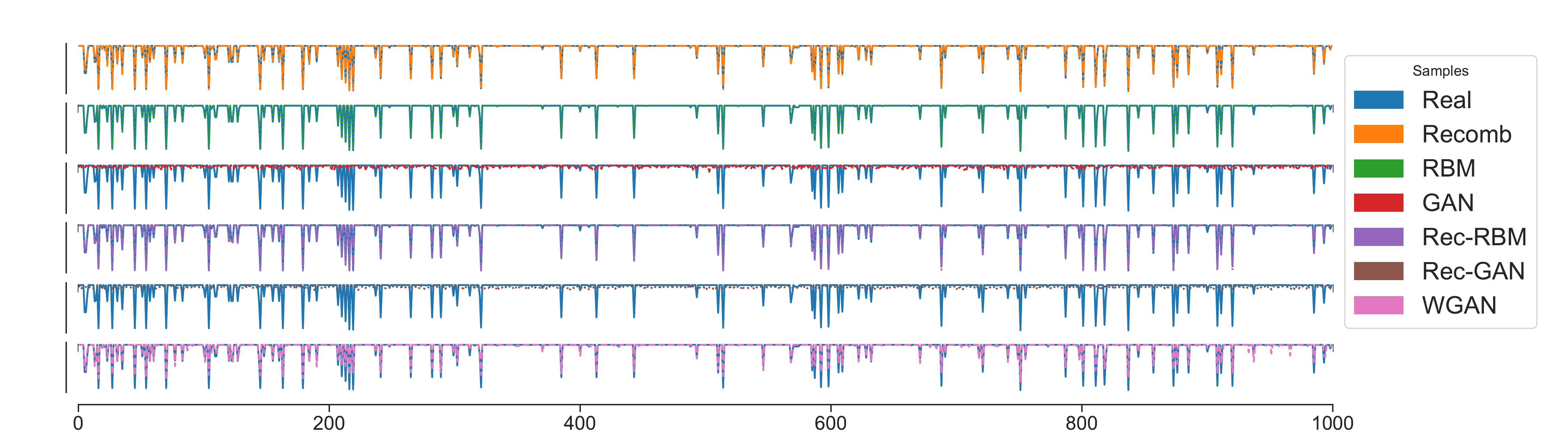}\label{fig:maf1000}}
\caption{Major allele frequencies for synthetic data generated by the models, plotted against the real data.} %
\label{fig:maf}
\vspace{-0.2cm}
\end{figure*}

\section{Utility Evaluation}\label{sec:utility}
This section presents a comprehensive utility evaluation of the synthetic data generated by the models introduced in Section~\ref{sec:models}.
We look at common summary statistics used in genome-wide association studies, aiming to assess the accuracy loss due to the use of synthetic datasets.  We choose summary statistics as our main utility evaluation metrics since prior work shows that they provide a good starting point for assessing the utility of synthetics genomic data~\cite{benner2017prospects,prive2021finding}.
While high(er) granularity may be required in some biomedical applications, if most generative models do not accurately replicate the high-level statistics of the real data, then it is extremely unlikely they would provide reasonable accuracy on single sequences.
More specifically, we analyze how well data generated by the generative models preserve allele frequencies, population statistics, and linkage disequilibrium.  We also compute the KS two-sample test~\cite{Hodges58} for the goodness of fit on the most relevant metrics in each subsection for each real dataset vs.~the synthetic data.

\subsection{Allele Statistics} \label{subsec:maf}

\begin{table}[t]
\medskip
\centering
\small
\begin{tabular}{@{}llrrrrr@{}}
\toprule
{\bf Models} & \multicolumn{2}{c}{\bf CEU} & \multicolumn{2}{c}{\bf CHB} & \multicolumn{2}{c}{\bf 1000 Geno} \\ %
              & $D$                      & p-value    & $D$  & p-value    & $D$  & p-value                 \\
\midrule
               \multicolumn{7}{c}{\bf MAF}                                                                \\
\midrule
{\bf Recomb}  & 0.04                   & 0.36       		& 0.03 & 0.53       		& 0.008 & 0.99                \\
{\bf RBM}     & 0.47                     & ${<}0.001$ 	& 0.33 & ${<}0.001$ & 0.06 & 0.03                    \\
{\bf GAN}     & 0.37                     & ${<}0.001$ 	& 0.52 & ${<}0.001$ & 0.58 & ${<}0.001$ \\
{\bf Rec-RBM} & 0.31                    & ${<}0.01$  	& 0.23 & ${<}0.01$  & 0.08 & ${<}0.1$                    \\
{\bf Rec-GAN} & 0.54                     & ${<}0.001$ & 0.61 & ${<}0.001$ & 0.863 & 0               \\
{\bf WGAN}    & 0.10                     & ${<}0.001$    & 0.14 & ${<}0.1$   & 0.28 & 0.82                \\\midrule
               \multicolumn{7}{c}{\bf SFS}                                                                \\
\midrule
{\bf Recomb}  & 0.07                     & 0.89       & 0.07 & 0.89       & 0.18 & ${<}0.1$                \\
{\bf RBM}     & 0.34                     & ${<}0.001$ & 0.28 & ${<}0.001$ & 0.12 & 0.44                    \\
{\bf GAN}     & 0.40                     & ${<}0.001$ & 0.41 & ${<}0.001$ & 0.23 & \hspace*{-1cm}${<}0.01$ \\
{\bf Rec-RBM} & 0.26                     & ${<}0.01$  & 0.23 & ${<}0.01$  & 0.06 & 0.99                    \\
{\bf Rec-GAN} & 0.64                     & ${<}0.001$ & 0.58 & ${<}0.001$ & 0.25 & ${<}0.01$               \\
{\bf WGAN}    & 0.09                     & 0.79       & 0.18 & ${<}0.1$   & 0.19 & ${<}0.1$                \\
\midrule
               \multicolumn{7}{c}{\bf \% Heterozygous Samples}                                            \\
\midrule
{\bf Recomb}  & \multicolumn{1}{r}{0.19} & 0.47       & 0.29 & ${<}0.001$ & 0.32 & ${<}0.001$              \\
{\bf RBM}     & \multicolumn{1}{r}{0.64} & ${<}0.001$ & 0.70 & ${<}0.001$ & 0.13 & 0.34                    \\
{\bf GAN}     & \multicolumn{1}{r}{0.90} & ${<}0.001$ & 1.00 & ${<}0.001$ & 0.57 & ${<}0.001$              \\
{\bf Rec-RBM} & \multicolumn{1}{r}{0.99} & ${<}0.001$ & 1.00 & ${<}0.001$ & 0.40 & ${<}0.001$              \\
{\bf Rec-GAN} & \multicolumn{1}{r}{0.55} & ${<}0.001$ & 0.68 & ${<}0.001$ & 0.52 & ${<}0.001$              \\
{\bf WGAN}    & \multicolumn{1}{r}{0.17} & ${<}0.1$   & 0.39 & ${<}0.001$ & 0.45 & ${<}0.001$ \\
\midrule
               \multicolumn{7}{c}{\bf Linkage Disequilibrium}                                            \\
\midrule
{\bf Recomb}  & \multicolumn{1}{r}{0.11} &  ${<}0.001$     & 0.12 & ${<}0.001$ & 0.31 & 0              \\
{\bf RBM}     & \multicolumn{1}{r}{0.61} & 0 & 0.69 & 0 & 0.08 & ${<}0.001$                 \\
{\bf GAN}     & \multicolumn{1}{r}{0.07} & ${<}0.001$ & 0.09 & ${<}0.001$ & 0.86 & 0            \\
{\bf Rec-RBM} & \multicolumn{1}{r}{0.40} & 0 & 0.41 & 0 & 0.38 & 0             \\
{\bf Rec-GAN} & \multicolumn{1}{r}{0.23} & 0 & 0.21 & 0 & 0.91 & 0              \\
{\bf WGAN}    & \multicolumn{1}{r}{0.10} & ${<}0.001$   & 0.09 & ${<}0.001$ & 0.53 & 0 \\ %
 \bottomrule
\end{tabular}%
\caption{Two-sample (real vs.~synthetic data) Kolmogorov-Smirnov test performed on the the main utility summary statistics presented in Section~\ref{sec:utility}.}
\label{table:ks}
\vspace{-0.3cm}
\end{table}

\descr{Major Allele Frequency (MAF).} In population genetics, the \emph{major allele frequency} (MAF) is routinely used to provide helpful information to differentiate between common and rare variants in the population, as it quantifies the frequency at which the most common allele occurs in a given population.
We start our utility analysis by comparing MAFs in the synthetic data vs.~the real data.

In Fig.~\ref{fig:maf}, we plot the MAF at each position for the real datasets and the synthetic samples, over the CEU and CHB populations, and the 1000 Genomes dataset.
For CEU/CHB (Fig.~\ref{fig:mafceu}--\ref{fig:mafchb}),the Recomb and WGAN replicate best the allele frequencies in the real data.
On the other hand, GAN and Rec-GAN fail to do so, and in fact, the generated samples seem random.
Even though not as close to the real frequencies as Recomb, the RBM model performs better than the GAN and Rec-GAN models.
In fact, RBM further improves when combined with Recomb (see Rec-RBM).

For 1000 Genomes (Fig.~\ref{fig:maf1000}),
Recomb's MAF distribution is also similar to the real data's.
However, RBM and Rec-RBM both display MAFs close to the real data, whereas, even with more training samples available, the GAN and Rec-GAN models still seem to produce random results.
Moreover, WGAN does not match the MAF distribution for this population as closely.
Overall, the difference in the MAF distributions across datasets is likely due to fewer samples available for the HapMap populations than the 1000 Genomes.
To further support our findings,  we also compute the KS two-sample test~\cite{Hodges58} on the statistics presented in this section (see Table~\ref{table:ks}).
The test compares the agreement between the cumulative distributions of two independent samples.
If the resulting $D$ value is low, and the p-value is high, we cannot reject the null hypothesis (i.e., there is no difference between the distributions).
For every two-samples test, the 95\% critical value is approximately 0.195 (as we have 100 samples in each dataset), so we can reject the null hypothesis (that there is no difference between the distributions) for all synthetic data above this value.
For the CEU and CHB populations, we cannot reject the null hypothesis for samples generated by the Recomb and WGAN., whereas for the 1000 Genomes dataset we cannot reject the null hypothesis for Recomb, RBM and Rec-RBM.

\descr{Alternate Allele Correlation (AAC).} To evaluate whether the real and synthetic data are {\em genetically} different, in Appendix Fig.~\ref{fig:aac}, we plot the \emph{alternate allele correlation} (AAC).
The more similar the two populations are, the closer the SNPs should be to the diagonal, as in the leftmost plots, where we have the real data against itself.
The strongest AAC is with the synthetic data generated by Recomb.
On the opposite side of the spectrum, the synthetic data generated by GAN and Rec-GAN have weak correlations.
For the CEU and CHB populations, %
Rec-RBM yields stronger AACs than simple RBM and the WGAN.
For the 1000 genomes dataset (Fig.~\ref{fig:aac1000}), there is a strong correlation between the alternate alleles for the real data and Recomb, RBM, Rec-RBM, and WGAN.

\descr{Site Frequency Spectrum (SFS).} Another summary statistic that captures essential information about the underlying distribution of the allele frequencies of a given set of SNPs in a population or sample is the SFS~\cite{Fisher31, EvansSS07}.
Basically, it provides a histogram whose size depends on the number of sequenced individuals.
In Fig.~\ref{fig:sfs}, we plot the scaled folded SFS, which is the distribution of counts of minor alleles in a sample calculated over all segregating sites.
We scale this value so that a constant value is expected across the spectrum for neutral variation and constant population size, which yields the best visual comparisons.
If the distribution of allele frequencies for the synthetic samples matches that of the real data, we would expect to see the two spectra aligned.

With the HapMap populations (Fig.~\ref{fig:sfsceu}--~\ref{fig:sfschb}), Rec-GAN suggests an excess of rare variants for a minor allele frequency around 0.1.
Whereas GAN seems to generate data closer to a neutral expectation, i.e., the synthetic dataset describes a more stable population.
Similarly, for the 1000 Genomes (Fig.~\ref{fig:sfs1000}), Rec-GAN has an excess of rare variants for a minor allele frequency less than 0.1, and this is also displayed, at a lower scale, by the GAN-generated data.
After computing the KS-two sample test, we cannot reject the null hypothesis for the samples generated by the Recomb and the WGAN models, for both CEU and CHB populations.
We reject the null hypothesis for the 1000 Genomes dataset for synthetic data generated by the GAN and Rec-GAN.

\begin{figure*}[t]
\centering
\subfloat[CEU Population]{\includegraphics[width=0.85\textwidth]{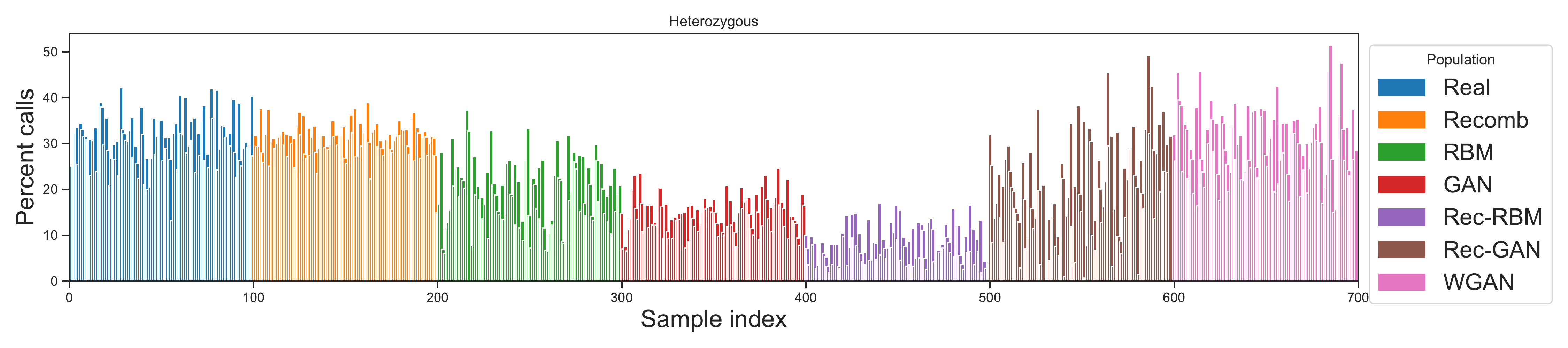}\label{fig:het-ceu}}\\[-0.15ex]
\subfloat[CHB Population]{\includegraphics[width=.85\textwidth]{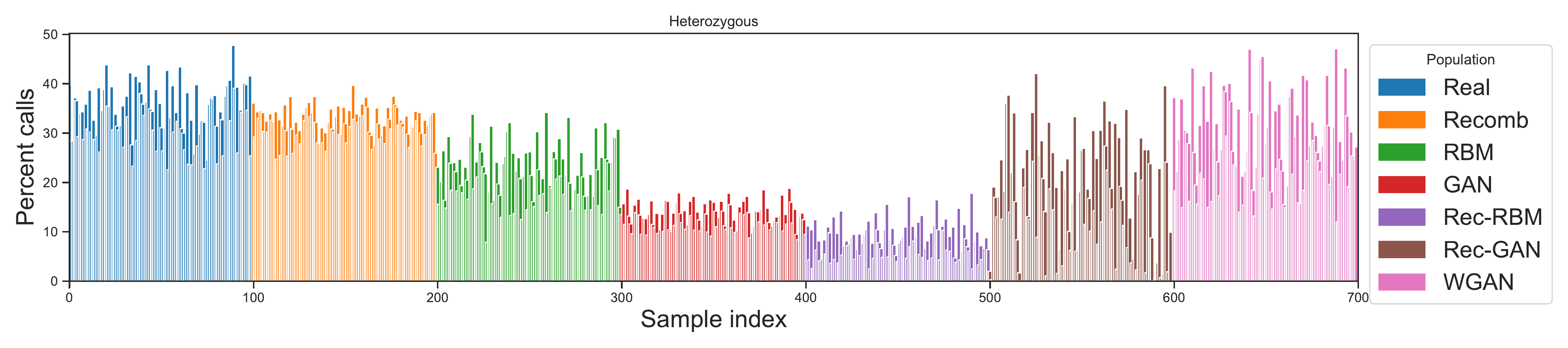}\label{fig:het-chb}}\\[-0.15ex]
\subfloat[1000 Genomes Population]{\includegraphics[width=.85\textwidth]{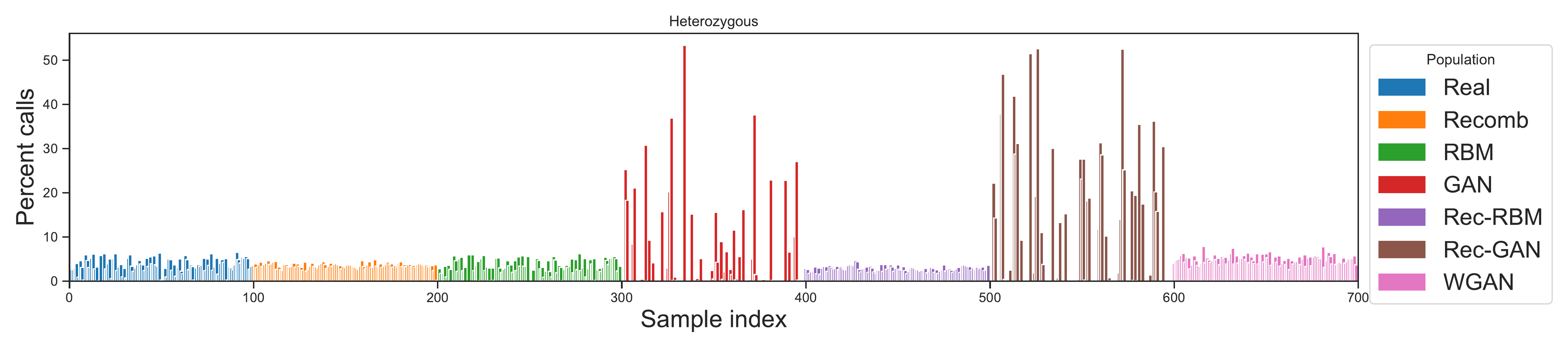}\label{fig:het_1000a}}\\[-0.15ex]
\subfloat[1000 Genomes Population]{\includegraphics[width=.85\textwidth]{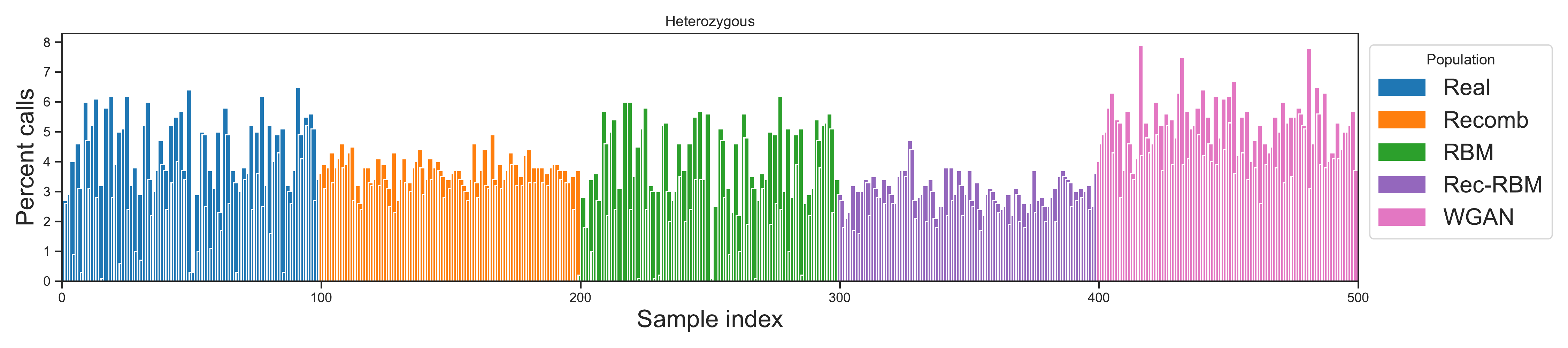}\label{fig:het_1000a_nogan}}\\
\caption{Percentage of heterozygous variants in each sample in the dataset for CEU, CHB populations, and the 1000 Genomes dataset.}
\label{fig:het}
\vspace{-0.3cm}
\end{figure*}

\subsection{Population Statistics}\label{subsec:pop_stat}

Next, we study population statistics to determine how close to the real dataset is the synthetic data.
In particular, we look at the percentage of heterozygous variants, the fixation index, and the Euclidean Genetic Distance.

\descr{Heterozygosity.} The condition of having two different alleles at a locus is denoted as heterozygosity.
The percentage of heterozygous variants is commonly used in population studies, as a low percentage of heterozygous variants implies less diversity in the population.
In Fig.~\ref{fig:het-ceu}--\ref{fig:het-chb}, we plot the percentage of heterozygous variants in each sample for the CEU/CHB populations, comparing the real statistics (blue/leftmost bars) vs.~those computed on the synthetic data.
From a visual inspection, both CEU and CHB datasets, the Recomb and WGAN samples yield a similar distribution to the real data. However, when computing the KS statistic, we find that we cannot reject the null hypothesis only for those two models for the CEU datasets, and we reject the null hypothesis for all synthetic datasets generated from the CHB dataset.

For the GAN and RBM, the percentage of heterozygous samples decreases, suggesting that both models produce more homozygous variants.
Moreover, even though for the major allele frequencies, Rec-RBM produces variants with statistics closer to the real data, the percentage of heterozygous variants turns out to be the lowest for both populations.
By contrast, Rec-GAN produces a higher percentage of heterozygous variants than GAN, even though the major allele frequencies are not aligned with the original samples.
With the 1000 Genomes (Fig.~\ref{fig:het_1000a}), the \% of heterozygous samples in the real data is lower across all samples.
Once again, and in line with previous results, GAN and the Rec-GAN significantly deviate from the \%s of heterozygous samples found in the real data.
In fact, in Fig.~\ref{fig:het_1000a_nogan}, we remove the GAN and the Rec-GAN models in order to take a closer look at the other models; with more data samples, the samples generated yield percentages of heterozygous samples relatively similar to the real data.  For this dataset, the only model for which we cannot reject the null hypothesis is the RBM.

\begin{figure}[t]
\centering
\includegraphics[width=0.85\columnwidth]{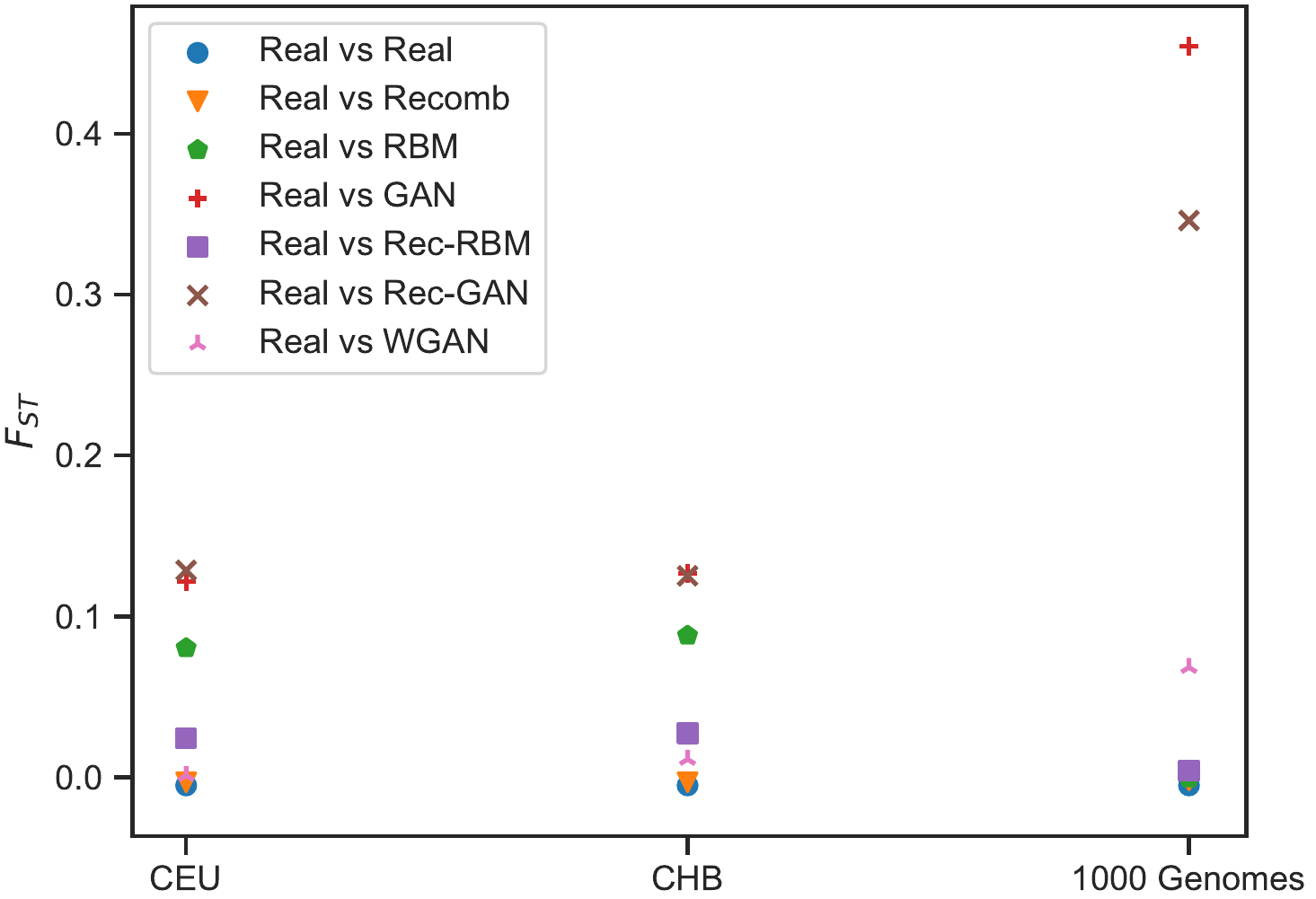}
\caption{Fixation index values ($F_{ST}$) for the CEU and CHB populations, and the 1000 Genomes dataset.}
\label{fig:fst}
\vspace{-0.2cm}
\end{figure}

\descr{Fixation Index ($F_{ST}$).} Another way to assess how {\em different} are groups of populations from each other is to use the fixation index \cite{HolsingerW09}.
It provides a comparison of differences in allele frequency, with values ranging from 0 (not different) to 1 (completely different/no alleles in common).
In Fig.~\ref{fig:fst}, we compare the $F_{ST}$ values for the real data against the synthetic samples.
For illustration purposes, we also include $F_{ST}$ of the real data against itself, which obviously yields $F_{ST}=0$.

Recomb is once again the closest to the real data, which confirms the alignment from Fig.~\ref{fig:maf} of the allele frequencies of the synthetic recombination data with the real data.
The $F_{ST}$ value for the synthetic data produced by RBM is, for both CEU and CHB populations, less than 0.10; however, the hybrid Rec-RBM model further reduces this value to less than 0.04, and so does WGAN.
For both populations, data generated by GAN and Rec-GAN has the highest $F_{ST}$, although, for the CHB population, the latter increases it, and for the CEU population reduces it.
Finally, for the 1000 genomes, Recomb, RBM, and Rec-RBM all have $F_{ST}$ close to the real data.
While still having a low $F_{ST}$, WGAN has a slightly higher value.
Whereas, with GAN and Rec-GAN, $F_{ST}$ significantly deviates from the real data, even with the increased number of samples of this dataset.

\descr{Euclidean Genetic Distance (EGD).} Since the fixation index does not easily allow for pairwise comparisons among populations, in Appendix Fig.~\ref{fig:ed}, we plot the Euclidean Genetic Distance (EGD) between the samples in each dataset.
EGD is routinely used as a measure of divergence between populations and shows the number of differences, or mutations, between two populations; a distance of $0$ means there is no difference in the results, i.e., there is an exact match.
From Fig.~\ref{fig:ed-ceu}--\ref{fig:ed-chb}, where the EGD on the diagonal is $0$, we observe that, for both CEU and CHB populations, the synthetic samples generated by GAN are closer to each other than by the other models.
Rec-GAN generates samples with EGD close to 0, suggesting that there are very few differences between them and samples with a distance of around 30.
As for the other population statistics, Recomb generates samples that match the differences observed in the real data the closest for both populations.
For RBM, the samples generated have fewer differences than the real data.
Perhaps more interestingly, Rec-RBM yields samples with a higher divergence than the real data; this can be a consequence of the low percentage of heterozygous samples found in the synthetic samples generated by this model (recall Fig.~\ref{fig:het}).
The samples from WGAN match some of the differences observed in the real data, but the model also yields a few samples with a higher divergence. %

\begin{figure*}[t]
\centering
\subfloat[CEU]{\includegraphics[width=0.99\textwidth]{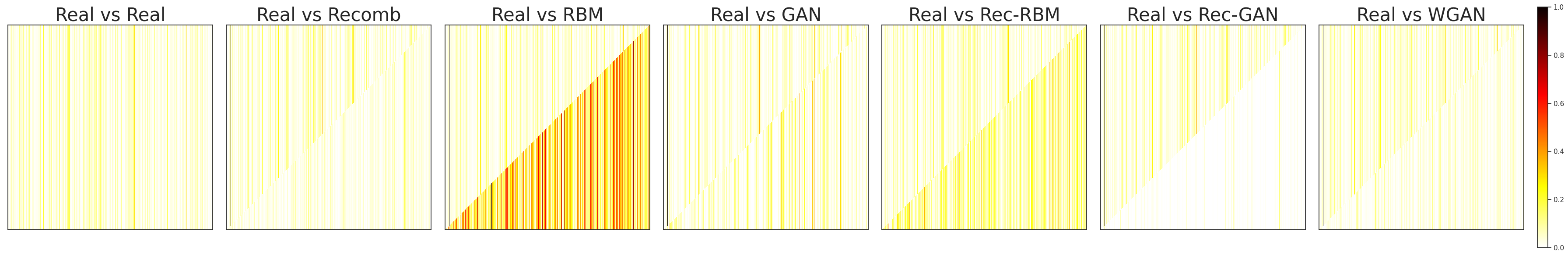}\label{fig:ld-ceu}}\vspace{-0.15cm}\\
\subfloat[CHB]{\includegraphics[width=0.99\textwidth]{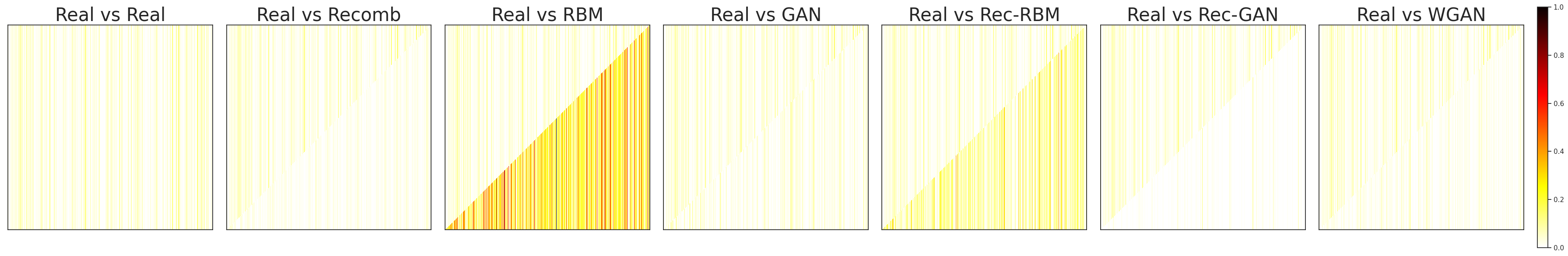}\label{fig:ld-chb}}\vspace{-0.2cm}\\
\subfloat[1000 Genomes]{\includegraphics[width=0.99\textwidth]{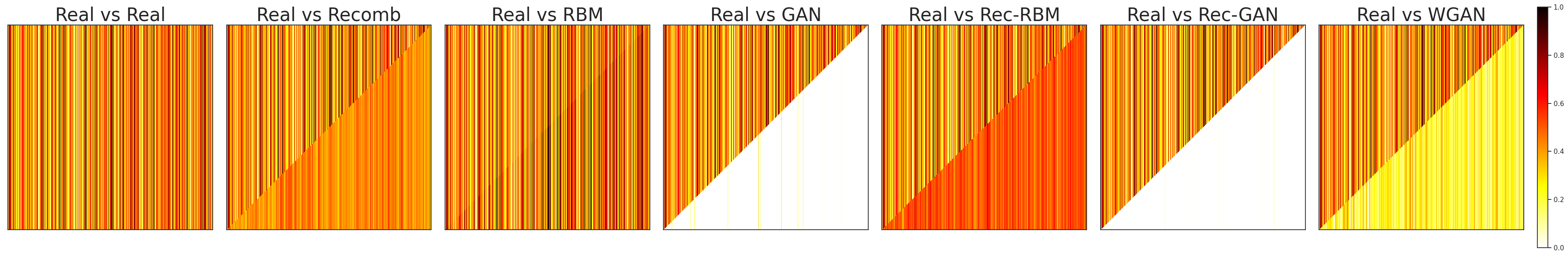}\label{fig:ld-1000}}
\caption{Pairwise Linkage Disequilibrium for Real vs.~Synthetic Samples.}
\label{fig:ld}
\vspace{-0.3cm}
\end{figure*}

Finally, for the 1000 Genomes (Fig.~\ref{fig:ed-1000}), all samples in the real data have closer EGDs between each other.
In fact, the samples generated by RBM yield a similar pattern in the EGD distances.
Although Recomb, Rec-RBM, and WGAN do too, they exhibit a lower distance, on average, between samples.
As for CEU/CHB populations, GAN and Rec-GAN models fail to capture the differences between samples.

\subsection{Linkage Disequilibrium (LD) Analysis}
Linkage disequilibrium (LD) captures the non-random association of alleles at two or more positions in a general population -- i.e., those alleles do not occur randomly with respect to each other.
In Genome-Wide Association Studies, LD allows researchers to optimize genetic studies, e.g., by preventing genotyping SNPs that provide redundant information~\cite{BushM12}.
In Fig.~\ref{fig:ld},
we plot the $r^2$ value for LD based on the Rogers-Huff method~\cite{RogersH09}.
This ranges from 0 (there is no LD between the 2 SNPs) to 1 (the SNPs are in complete LD, i.e., the two SNPs have not been separated by recombination and have the same allele frequencies).

For CEU and CHB populations, RBM generates samples that display a stronger LD than the real data.
With more training samples, Rec-RBM yields a weaker LD but still stronger than the real data for both models.
On the other side of the spectrum, for Rec-GAN, the LD for the synthetic data is the weakest. %
For the 1000 Genomes, we find a stronger LD between the real samples than with the other two datasets.
RBM generates samples that are almost indistinguishable from the real data in terms of LD.
The LD in the synthetic datasets generated by Recomb, Rec-RBM, and WGAN have lower correlations than RBM, with GAN and Rec-GAN both failing to preserve the LD. %
In this case, the non-parametric KS test yields more surprising results for both the CEU and CHB populations. We find that we cannot reject the null hypothesis for samples generated by Recomb, GAN and WGAN.  This implies that, even though the GAN model does not preserve well the allele frequencies or the population statistics, it manages to preserve the correlations between alleles.

For the 1000 Genomes data, the only generative models that yields samples which cannot be rejected under the null hypothesis is the RBM.

\subsection{Takeaways}
Our utility evaluation shows that %
there are only a handful of cases where generative models produce synthetic genomic data with high utility on popular tasks.

The Recomb model, which is based on high-order SNV correlations, generates synthetic data preserving most statistical properties displayed by the real data, even when few samples are available.
We get better utility when the genetic map is included with the data rather than generated from the existing data.
This conclusion is also supported by the fact that we cannot reject the null hypothesis for this model for most statistics analyzed (with the exception of the percentage of heterozygous samples for the CHB).

With RBM, more training samples improve the quality of the synthetic data, as evidenced by the difference between the HapMap populations and the 1000 Genomes dataset.

We also find that when few samples are available for training, the hybrid Rec-RBM model approach helps improve the quality of samples compared to just RBM, but the samples are not as close to the real data as for Recomb.
This is clear from the utility of the synthetic data on the two smaller HapMap datasets.
For the 1000 genomes, it is not surprising that Rec-RBM's performance is worse than RBM since Recomb does not generate as ``useful'' samples as for the other two datasets.
Finally, the GAN and the Rec-GAN models generate samples with the lowest utility, regardless of the number of samples available for training. However, even though the allele and population statistics are now well preserved, our non-parametric test showcases that the correlations between alleles are preserved for the LD.
In contrast to the other GAN models studied, the data generated by WGAN preserves most statistical properties of the real data.

Overall, our analysis exposes the limitations, in terms of statistical utility, of using generative machine learning models to produce synthetic genomic data.

\section{Privacy Evaluation}\label{sec:privacy}

Next, we quantify the privacy provided by synthetic data by evaluating its vulnerability to Membership Inference Attacks (MIAs).
More precisely, we compute the Privacy Gain, PG, obtained by releasing a synthetic dataset instead of the real data (see Section~\ref{subsec:pg}). %
Recall that if the synthetic data does not hide nor give any additional information to an MIA attacker, $PG_t$, for a target record $t$, should have a value of around 0.25.

We present experiments for both a ``standard'' MIA and a novel attack, which we denote as {\em MIA with partial information.}
The latter essentially assumes that the adversary only has access to partial data from the target sequence.
We exclude GAN and Rec-GAN from the evaluation since they yield poor utility performance, so there is not really any point in evaluating their privacy.

Throughout our evaluation, we randomly choose 10 targets from each dataset across 10 test runs.
In each run, we fix the target and sample a new training cohort.
We train the attack classifier using 5 shadow models, using 100 synthetic training sets for each of them.
We then compute the PG on 100 synthetic datasets, with a split of 50 sets generated from a training set including the target and 50 sets generated without.
Finally, we report the PG for each test and each target as the average PG across all synthetic datasets tested.

\subsection{Privacy Gain Under Membership Inference Attack} \label{sec:pg_mia}

\descr{Threat Model.} We assume that the adversary has access to the target, to the synthetic datasets,  as well as public datasets with similar distribution to the original data (which may or may not include the target) for training his shadow models. The goal of the adversary is to identify whether the target sequence was used for training the model that generated the synthetic dataset.  This could lead to the leakage of additional information about the target, as for example if a synthetic dataset of cancer patients is released and the adversary can identify that the target is part of the training set, it would lead to exposing the cancer status of the target.

We use three adversarial classifiers: K-Nearest Neighbor (KNN), Logistic Regression (LogReg), and Random Forest (RandForest).
We use four feature sets, as described in Section~\ref{subsec:pg}: Naive (\tikzcircle[colorFnaive, fill=colorFnaive]{2pt}$\Fnaive$), Histogram (\tikzcircle[colorFhist, fill=colorFhist]{2pt}$\Fhist$ ), Correlations (\tikzcircle[colorFcorr, fill=colorFcorr]{2pt}$\Fcorr$), and an Ensemble feature set (\tikzcircle[colorFens, fill=colorFens]{2pt}$\Fens$).

\begin{figure*}[t]
\centering
\includegraphics[width=0.25\textwidth]{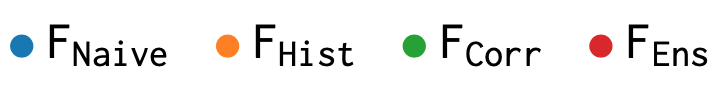}\\[-0.5ex]
\subfloat[CEU Population]{\includegraphics[width=0.9\textwidth]{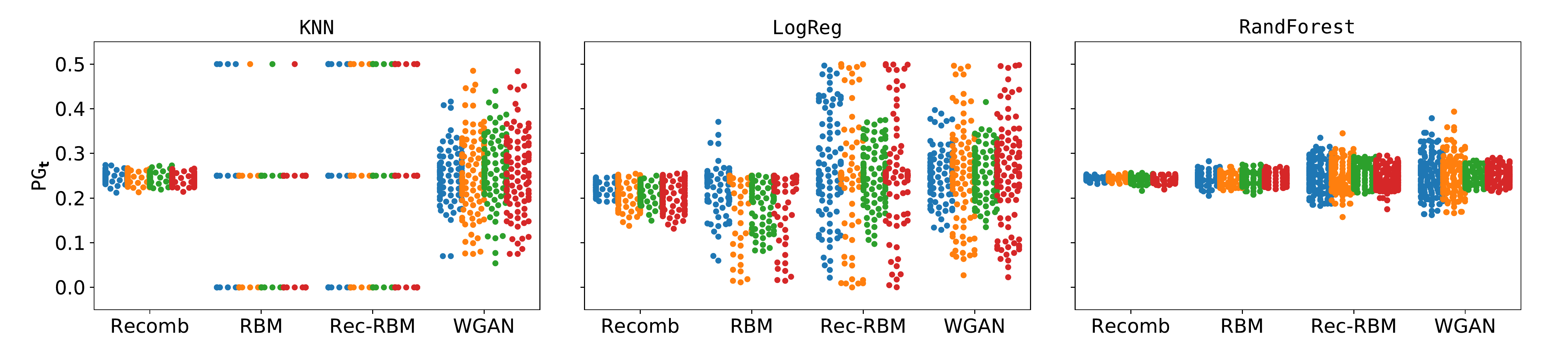}\label{fig:mia_ceu}}\\[-0.25ex]
\subfloat[CHB Population]{\includegraphics[width=0.9\textwidth]{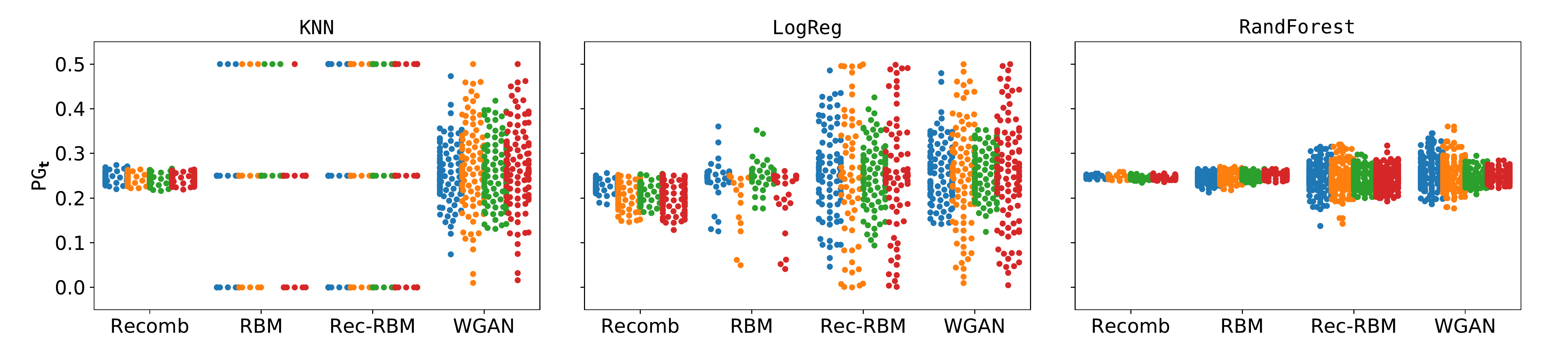}\label{fig:mia_chb}}
\caption{Privacy Gain (PG) of different models over the two HapMap populations.} %
\label{fig:mia}
\vspace{-0.3cm}
\end{figure*}

\subsubsection{HapMap Populations}

In Fig.~\ref{fig:mia}, we report the PG value for targets randomly chosen from the two HapMap populations.

\descr{KNN.} For CEU, using KNN (Fig.~\ref{fig:mia_ceu}, left), we find that over 74\% of the targets in the synthetic dataset generated by Recomb have a PG lower than the random baseline (0.25) for the Ensemble, Correlations, and Histogram feature sets.
With RBM, there are between 84\% and 88\% of the targets, depending on the feature set, that have a PG of 0.25; in other words, for these targets, the adversary's probability of inferring their presence in the training set is the same as random guessing.
However, between 10\% and 15\% of the targets have no PG at all, whereas there are between 1\% and 4\% of the targets, depending on the feature set, for which the synthetic data perfectly protects the target from MIA (PG=0.5).
With Rec-RBM, at least 59\% of the targets have a PG of 0.25 under the four feature sets.
With WGAN, at least 48\% of the targets have a PG of 0.25, depending on the feature set.

For the CHB population (Fig.~\ref{fig:mia_chb}, left), we find that over 60\% of the targets generated by the Recomb, with all features, have PG below the random guess baseline.
With RBM, between 89\% and 97\% of targets have PG of exactly 0.25 across all feature sets, i.e., the synthetic dataset generated by the RBM for these targets does not hide nor give new information to the attacker about their membership to the synthetic dataset.
Interestingly, for the Correlations feature set, no target has PG lower than 0.25.
As for the CEU population, %
about 50\% of the targets across all feature sets have a PG of 0.25 for data from Rec-RBM and at least 47\% of targets from WGAN.

\begin{figure*}[t]
\begin{center}\includegraphics[width=0.25\textwidth]{graphs/plots/legend.png}\end{center}\vspace*{-0.75cm} %
\subfloat[Random Target]{\includegraphics[width=0.4\textwidth]{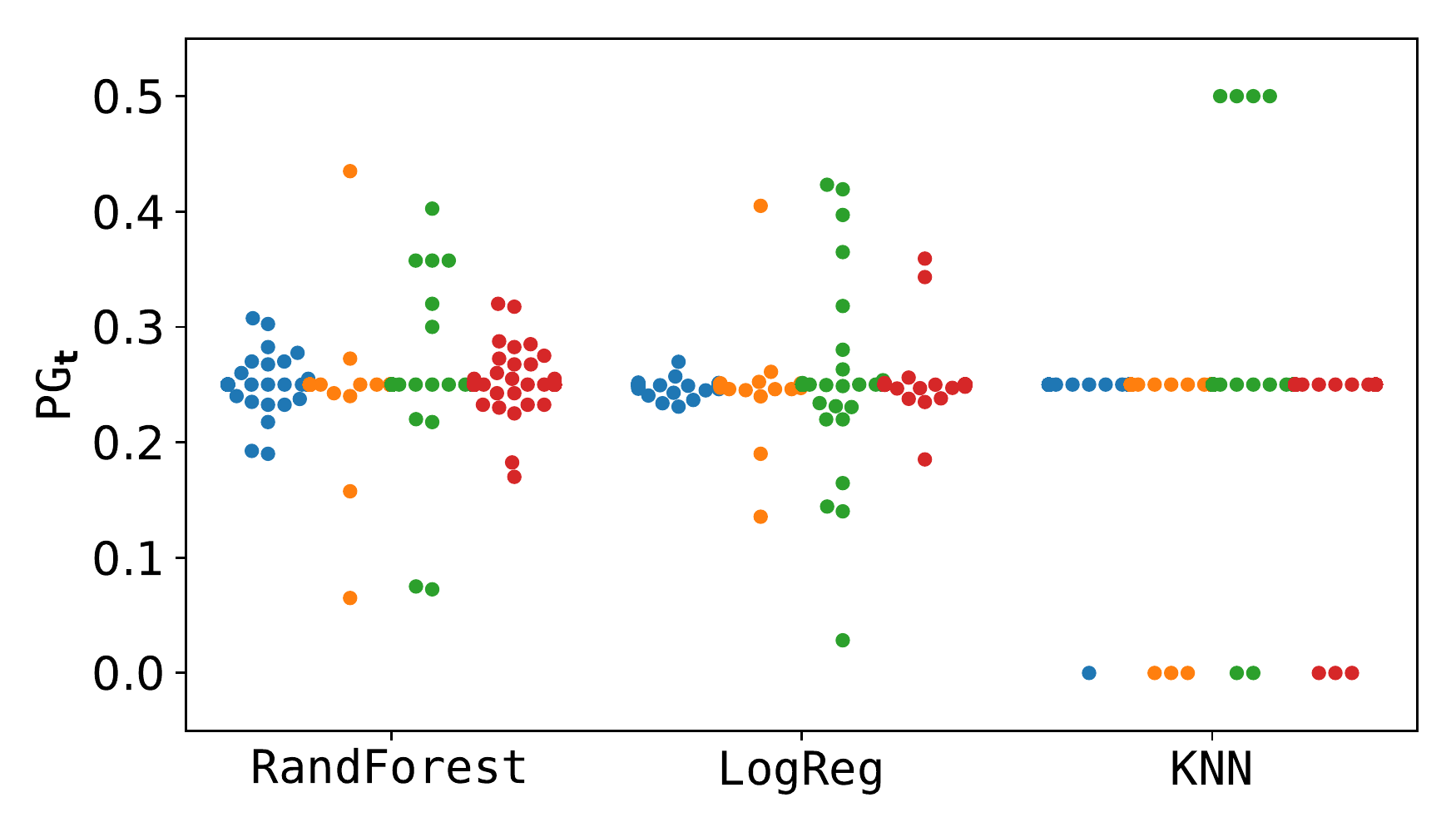}\label{fig:mia_1000}}
\hspace{0.3cm}
\centering
\subfloat[Outlier target]{\includegraphics[width=0.4\textwidth]{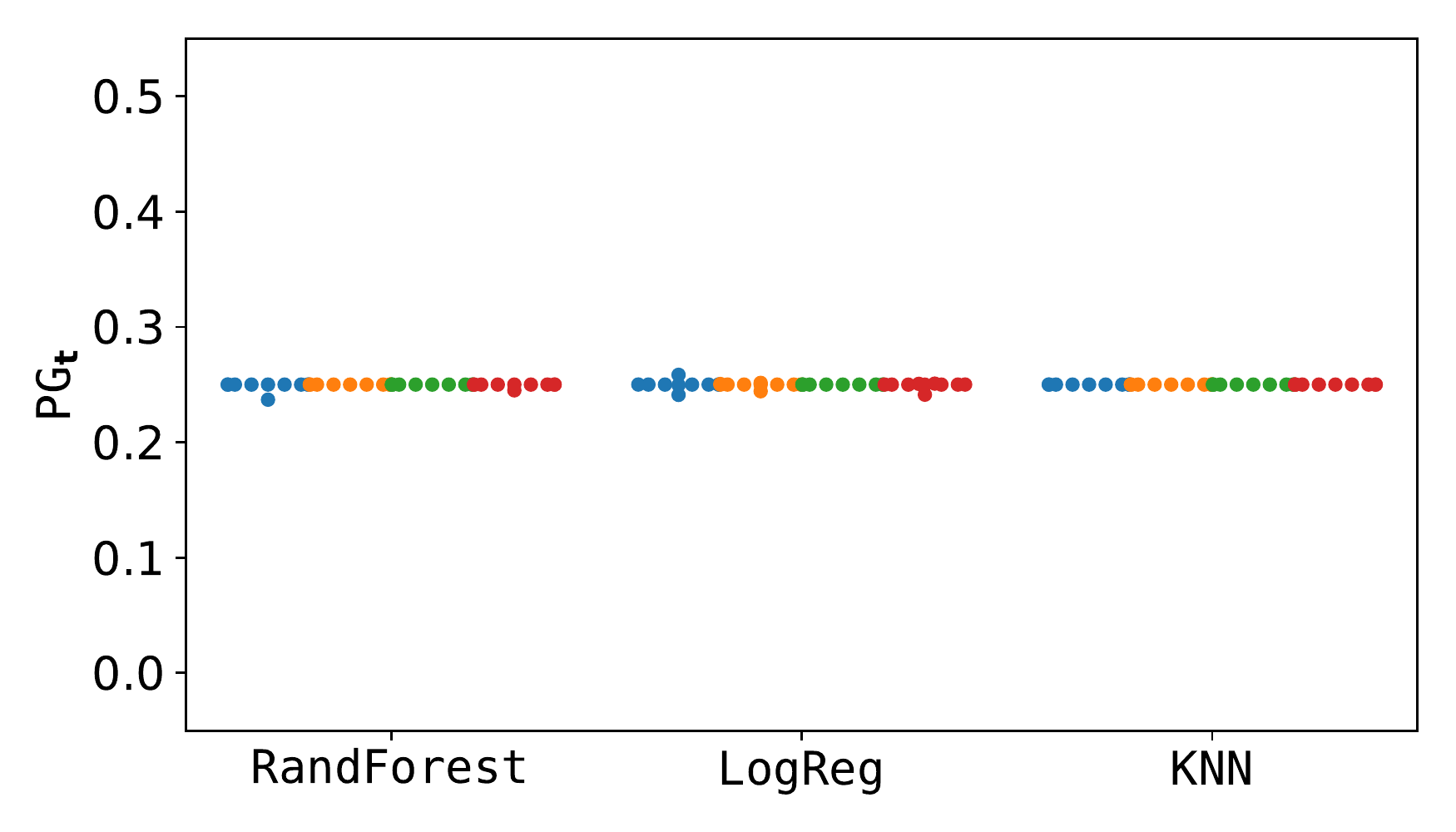}\label{fig:mia_out}}
\caption{Privacy Gain (PG) of RBM over the 1000 Genomes dataset. }
\label{fig:mia1000}
\vspace{-0.5cm}\end{figure*}

\descr{LogReg.} Using LogReg, both Recomb and RBM have the lowest PG among all attack classifiers for both HapMap populations.
For CEU (Fig.~\ref{fig:mia_ceu}, middle), using the Histogram feature set, 94\% (resp., 96\%) of the targets from Recomb (resp., RBM) have PG below 0.25, which is the random guess baseline.
Under Correlations, 99\% (resp., 97\%) of the targets in Recomb (resp., RBM) have PG below 0.25,
while, for the Ensemble feature set, 96\% (resp., 98\%) of the targets from Recomb (resp., RBM) have PG below 0.25. %
With Rec-RBM, we find that between 52\% and 56\% of the targets across all feature sets have PG above 0.25, and with WGAN, between 50\% and 57\% of the targets across all feature sets have PG above 0.25.
Moreover, for the Rec-RBM and WGAN-generated data, no target consistently has a lower PG than the random guess baseline across all test runs.

For CHB (Fig.~\ref{fig:mia_chb}, middle), with synthetic data generated by Recomb, the average PG is below the random baseline (0.25) for 99\% of the targets in the Histogram feature set, 97\% for Correlations, and 96\% for Ensemble. %
For RBM, 79\% of the targets in the Histogram feature set have a PG below 0.25. %
Under the Ensemble feature set, 84\% of the targets have PG below 0.25. %
For synthetic data generated by Rec-RBM, we find that 54\% of the targets from the Histogram and Ensemble feature sets have PG lower than 0.25, and 46\% have PG over 0.25.
For the Naive and Correlations feature sets, respectively, 45\% and 47\% of the targets have PG lower than the random guess baseline.
For WGAN-generated data, the Correlations feature set yields most targets (55\%) with PG$<$0.25.

\descr{RandForest.} When using RandForest as the attack classifier on data from the CEU population (Fig.~\ref{fig:mia_ceu}, right), with RBM-generated data, 73\% of the targets from both Correlation and Histogram feature sets have lower PG than the random baseline.
This is for 69\% and 62\% of the targets with, respectively, Ensemble and Naive feature sets.
For the synthetic data generated by Rec-RBM, about 51\% of targets have a PG of over 0.25, and 49\% of the targets have PG of less than 0.25 across all feature sets.
For WGAN, between 46\% and 59\% of the targets from all feature sets have a PG less than the random guess baseline (0.25), with the Correlations feature set having the least percentage of vulnerable targets (59\%).

For CHB (Fig.~\ref{fig:mia_chb}, right), %
the lowest privacy gain for the samples generated by Recomb: over 79\% of all targets for each of the four feature sets have PG lower than the random baseline.
For the synthetic samples from RBM, 71\%, 61\%, and 53\% of the targets under the Naive, Histogram, respectively, Ensemble feature sets have PG lower than 0.25.
However, for the Correlations feature sets, we find that 55\% of the targets have a PG of 0.25, meaning that those targets are protected from MIA.
For the synthetic samples generated by Rec-RBM, 54\% of the targets from the Histogram and Ensemble feature sets and 47 and 45\% of the targets from the Correlations and Naive, respectively, have PG lower than 0.25.
Finally, for the WGAN, we find that between 44\% and 53\% of the targets have PG$>$0.25 across all four feature sets.
To give a better idea of different trade-offs between utility and privacy, we also plot the PG vs.~the $F_{ST}$ (see Section~\ref{subsec:pop_stat}) for all the models presented in Appendix~\ref{app:tradeoff}.

\subsubsection{1000 Genomes Population} %

For this dataset, %
we focus our analysis on the RBM model, as it generated synthetic data closest to the real data across all utility metrics evaluated.

\descr{Random Target.} In Figure~\ref{fig:mia_1000}, we plot the PG for the synthetic data generated by RBM for randomly chosen targets.
With the RandForest MIA classifier, we observe that 56\%, 75\%, 92\%, and 50\% of the targets have PG higher than the random guess baseline (PG$\geq$0.25) for, respectively, Naive, Histogram, Correlations, and Ensemble feature sets.
The high percentage of targets that have a PG$\geq$0.25 for the Correlations suggests that the impact of a single target in the training dataset of the RBM on the correlations of the synthetic dataset is minimal.

Then, with the LogReg classifier, we find a high variation in PG, similar to the HapMap populations in the case of Rec-RBM.
Under the Naive feature set, half of the targets have PG gain below the random guess baseline.
Similarly, 43\%, 45\%, and 43\% of targets have a PG lower than 0.25 for, respectively, the Correlations, Histogram, and Ensemble feature sets.
Finally, for the KNN classifier, over 94\% of the targets have a PG of 0.25 across all feature sets.

\descr{Outlier Target.} To better understand whether or not, with more training data, a target's signal in the synthetic dataset is diluted, we also test an ``extreme'' outlier case.
That is, we craft an outlier target that has only minor alleles at all positions.
While we are aware that this case would be sporadic in a real-world scenario, our goal is to observe whether and how much this impacts PG.
To this end, in Figure~\ref{fig:mia_out}, we plot the PG of this outlier case across 10 test runs.

With RandForest, we find that PG is below 0.25 for 8 of the 10 test runs under the Ensemble feature set.
In fact, this is the only combination between attack classifier and feature set for which a greater percentage of the targets have a lower privacy gain than in the random target case.
For the Naive feature set, PG is below the random baseline in only 3 of the test runs. %
For the Correlations and Histogram feature sets, all test runs yield PG of 0.25 or above.
With LogReg, 4 out of 10 of the test runs for the Naive, Histogram, and Correlations feature sets yield PG below 0.25. %
For Ensemble, this happens for 6 test runs. %
Finally, with KNN, across all feature sets, PG for all test runs is 0.25, i.e., the synthetic data does not disclose any membership information regarding the outlier.

While there are differences across classifiers/feature sets, the PG is centered around 0.25 for all test runs in the outlier target setting.
This evident from Figure~\ref{fig:mia_out}, which implies that, across all test runs, the accuracy of the MIA is not much better than random guessing.
\subsubsection{Takeaways}
The different combinations of datasets, attack classifiers, and feature sets yield varied results with respect to privacy.
This is due to two main reasons: first, not all classifiers have the same accuracy on tasks for the same dataset, as shown in previous work~\cite{Amancio14}.
Second, the features that the generative model preserves after training will ``reflect'' in the synthetic data; thus, this will impact the PG based on the feature extraction method.
On the HapMap populations, while the utility experiments show that Recomb-generated synthetic data is ``closest'' to the real data, it does so with a significant privacy loss. %
The RBM-generated synthetic data is the most vulnerable under the LogReg classifier, with at least 70\% of the targets across both populations and all feature sets having PG below the random guess baseline.
This suggests that, with few data samples available for training, the RBM model is likely to overfit and is thus susceptible to MIAs.

For Rec-RBM and WGAN, the attacker cannot reliably predict membership, i.e., extra samples from the Recomb model in the training of the Rec-RBM dilute the target's signal in the training data.
However, for both models, we still find combinations of targets and training sets for the attack classifier for which PG is significantly lower than the random guess baseline; i.e., the synthetic data will still expose membership information about the respective targets.
On the 1000 Genomes, %
PG values have a higher variation overall when the target is chosen randomly from the dataset than for the two smaller HapMap datasets.
The results for RBM confirm our hypothesis that the target's influence is diluted within larger datasets. %
Once again, %
this does not mean that MIAs are not possible for Rec-RBM and WGAN; it depends on the combination of the target, training set, attack classifier, and feature set.

However, with an ``extreme'' outlier (i.e., a target with minor alleles at all positions), the synthetic data generated by RBM does not have a big impact on PG.
In this case, the PG is actually close to the random guess baseline across all test runs.

We find that targets get very different levels of privacy protection, based on the combination of target and training set used to generate the synthetic data.
For a privacy mechanism to provide good privacy protection, it needs to be predictable, i.e., all targets should have a privacy gain above the random guess baseline, which our evaluation shows it is not the case for the data generated by the models we have evaluated.

\subsection{Privacy Gain under Membership Inference Attack with Partial Information}\label{sec:new}

Next, we introduce a novel attack, which we denote as MIA with Partial Information (MIA-PI).
Basically, we only give the attacker access to a fraction of SNVs from the target sequence, chosen at random.
The attacker then uses the Recombination model from~\cite{SamaniHAEFH15} as an inference method to predict the rest of the sequence.
Compared to the previous attack, the adversary trains their (attack) classifier using the sequence {\em inferred} from the partial data.
Thus, the PG formula also needs to be adjusted to account for how likely an adversary is to identify a target %
from partial information.

\descr{Threat Model.} We assume that,  similar to the MIA attack, the adversary has access to the synthetic dataset, as well as public datasets with similar distribution to the original data (which may or may not include the target) for training his shadow models.  In contrast with the previous MIA, the adversary does not obtain access to the full sequence of the target, but only to a percentage of the SNPs from the target sequence (as for example, the victim undertakes some specific genomic tests and the adversary learns that information about the victim from the outcome of those tests), as well as public knowledge about genomic data.  The goal of the adversary is to identify if the partial target sequence was part of the training set which generated the synthetic data.

\descr{PG for MIA-PI.} Assuming the attacker has partial information $t'$ as a fraction of the SNVs from $t$, they first use the Recomb model, as an inference algorithm, to predict the rest of the SNVs from the target sequence, which we denote by $t_p$.
The privacy gain is computed as:%

{\footnotesize
$$ PG_{t}  = \frac{ \overline{MIA_{t_p}}(R_{t})- \overline{MIA_{t_p}}(S_{test})}{2} \mbox{, where} \vspace*{-0.1cm}$$
$$\overline{MIA_{t_p}}(S_{test}) = \sum_{S_i\in S_{test}} \frac{\Pr[MIA_{t_p}(S_i) = 1]}{2*n_s}\mbox{, and} \vspace*{-0.1cm}$$
$$\overline{MIA_{t_p}}(R_{t}) = \sum_{R_i \in R_{t}}\frac{\Pr[MIA_{t_p}(R_i) = 1]}{2*n_s}.$$
}

\noindent That is, the privacy gain, in this case, is computed as the difference between the probability that the attacker, who has partial information about the target record, correctly identifies that the target is part of the real dataset versus the target being part of the training set used to generate the synthetic dataset.

As a result, PG now ranges between -0.5 and 0.5, where 0.5 means that having the real dataset $R$ and the partial information $t'$ about the target allows the adversary to infer the membership of $t$ in $R$, while the synthetic dataset reduces the adversary's chance of success (i.e. $\overline{MIA_{t_p}}(R_{t})=1$ and $\overline{MIA_{t_p}}(S_{test}) = 0$).
A negative PG value means that publishing the synthetic data, instead of the real data, improves the adversary's chance to correctly infer membership of the target $t$ (i.e. $\overline{MIA_{t_p}}(R_{t})<\overline{MIA_{t_p}}(S_{test})$).
If publishing the synthetic data does not increase nor decrease the adversary's inference powers, we should have PG=0 (i.e. $\overline{MIA_{t_p}}(R_{t})=\overline{MIA_{t_p}}(S_{test})$).

\descr{Experiments.} From the experiments presented in Section~\ref{sec:pg_mia}, we find that the attack classifier that yields the lowest PG is Logistic Regression; thus, we only experiment with that one to ease presentation.
In the following, we present the results of the MIA-PI experiments for the {\bf\em CEU population}, focusing on the {\bf\em Recomb and RBM models} (as mentioned, with a LogReg attack classifier).
We do so as these two models yield the lowest PG in Section~\ref{sec:pg_mia}.

\begin{figure}[t]
\centering
{\includegraphics[width=0.35\columnwidth]{graphs/plots/legend.png}}\\[-0.5ex]
\includegraphics[width=1\columnwidth]{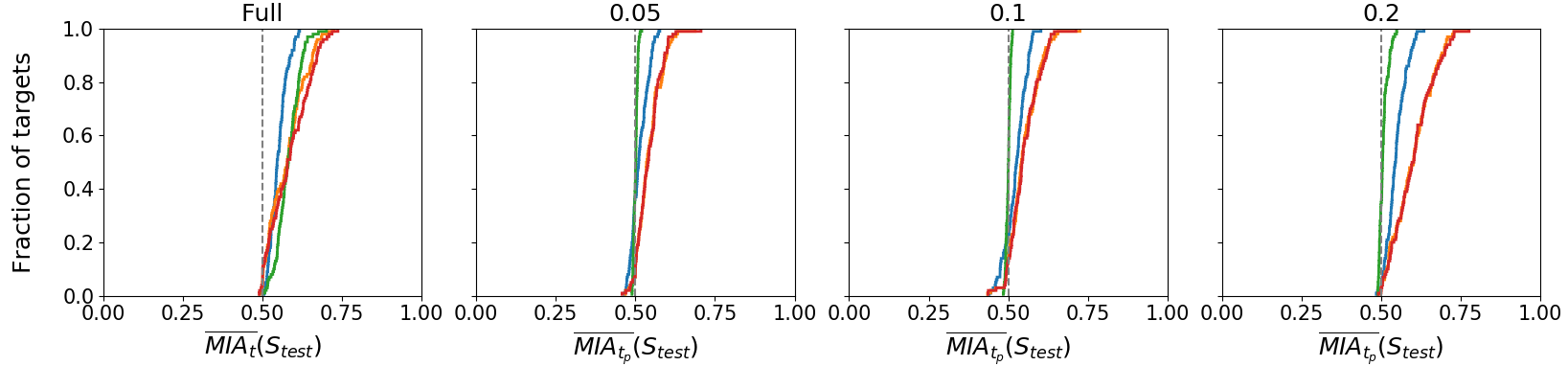}\hspace{-0.25cm}
\caption{Accuracy of the Membership Inference Attack with access to full and partial information (0.05, 0.1, and 0.2 ratio) for {\bf \em Recomb}.}
\label{fig:mia_acc_part_recomb_ceu}
\vspace{-0.25cm}
\end{figure}

\descr{Recomb.} In Figure~\ref{fig:mia_acc_part_recomb_ceu}, we plot the Cumulative Distribution Function (CDF) of the accuracy of the attack for Recomb when the adversary has access to the full sequence vs.~partial information, specifically, a ratio of 0.05, 0.1, and 0.2 of the total SNVs from the target sequence.
Interestingly, even when only 0.05 of the target SNVs are available to the attacker, for 90\% and 91\% of the targets from the Histogram and respectively Ensemble feature sets, the attacker's accuracy is still above the random guess baseline (50\% accuracy).
Our intuition is that many targets are vulnerable to the attack, even with little partial information, since we use the Recomb model not only for the attack but also as an inference method to predict the rest of the sequence.

\begin{figure}[t]
\centering
{\includegraphics[width=0.35\columnwidth]{graphs/plots/legend.png}}\\[-0.5ex]
\includegraphics[width=1\columnwidth]{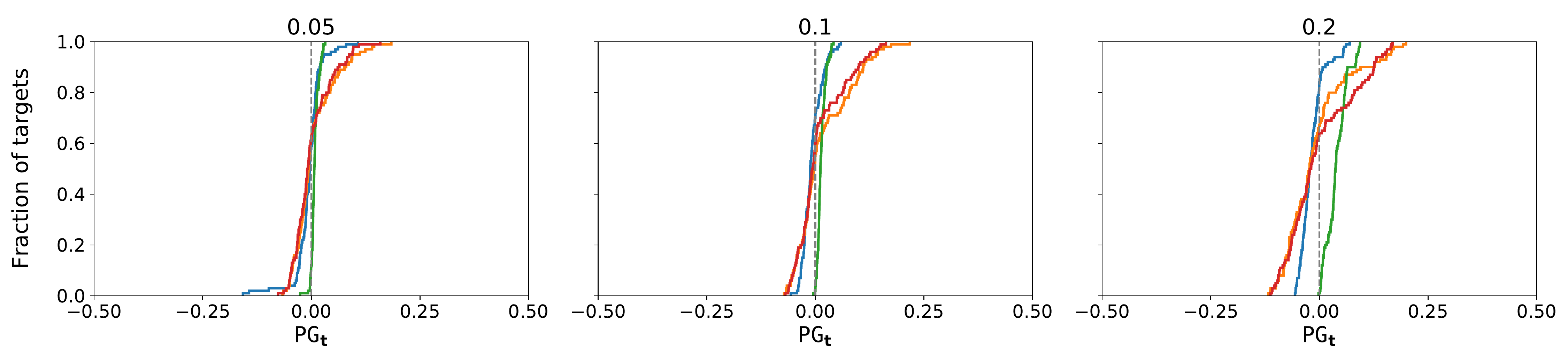}\hspace{-0.25cm}
\caption{Privacy Gain (PG) for synthetic samples from {\bf \em Recomb}.}
\label{fig:mia_part_recomb_ceu-pg}
\vspace{-0.2cm}
\end{figure}

To explore how much of the MIA-PI vulnerability is due to the release of synthetic datasets, and not only by how much information the attacker has available, in Figure~\ref{fig:mia_part_recomb_ceu-pg}, we plot the CDF of the PG with MIA-PI.
In line with the accuracy results, we find that the PG is greater than 0 for at least 88\% of the targets for all ratios of partial information tested in the case of the Correlations feature set.
However, for the other three feature sets, releasing the synthetic dataset instead of the real data decreases the privacy gain (i.e., PG<0) for the majority of targets.
When the adversary has access to just 5\% of the SNVs from the target, there is a negative PG for 61\% of the targets under the Histogram and 62\% of the targets under the Ensemble feature sets.
With 10\% of the target sequence available, 54\% and 60\% of the targets under, respectively, the Histogram and the Ensemble feature sets have negative PG, and with 20\%, these numbers go up to 64\% and 67\%.
There are more targets with negative PG with increasing partial information available to the attacker for the Naive feature set, i.e., 59\%, 70\%, and 84\% with, respectively, 5\%, 10\%, and 20\% of the target sequence available.
Overall, this shows that releasing the synthetic dataset instead of the synthetic data does not mitigate privacy, even when the attacker does not have access to the full sequence.

\begin{figure}[t]
\centering
{\includegraphics[width=0.35\columnwidth]{graphs/plots/legend.png}}\\[-0.5ex]
\includegraphics[width=1\columnwidth]{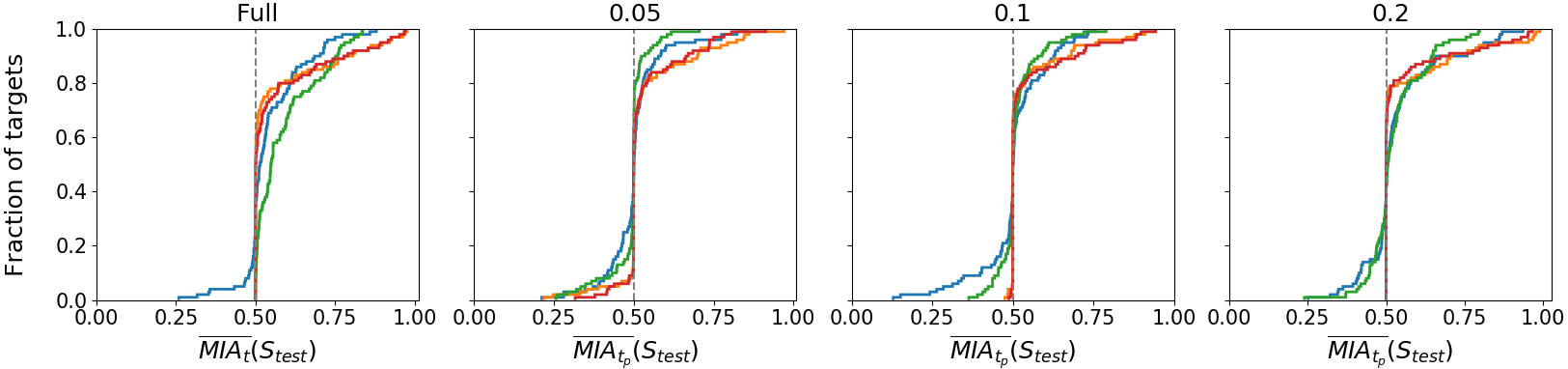}
\caption{Accuracy of the Membership Inference Attack with access to full and partial information (0.05, 0.1, and 0.2 ratio) for {\bf \em RBM}.}
\label{fig:mia_acc_part_rbm_ceu}
\vspace{-0.2cm}
\end{figure}

\begin{figure}[t]\centering
{\includegraphics[width=0.35\columnwidth]{graphs/plots/legend.png}}\\[-0.5ex]
\includegraphics[width=1\columnwidth]{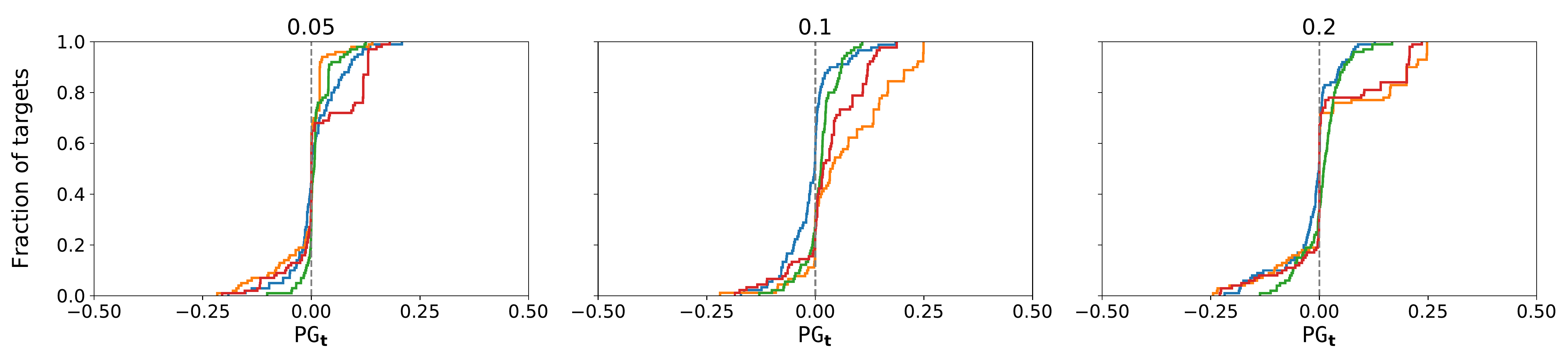}%
\caption{Privacy Gain (PG) for synthetic samples generated by {\bf \em RBM}.}
\label{fig:mia_part_rbm_ceu-pg}
\vspace{-0.2cm}
\end{figure}

\descr{RBM.} In Figure~\ref{fig:mia_acc_part_rbm_ceu}, we plot the CDF for the accuracy of the attack for the RBM model for both full and partial information about the target record available to the attacker.
Across all feature sets, there is an increase in the accuracy of the attack with more information available to the attacker, as is expected.
We also look at CDF for the PG in the case of partial information available to the attacker in Figure~\ref{fig:mia_part_rbm_ceu-pg}.
Once again, under the Naive feature set, increasing the partial information available to the attacker negatively correlates with the percentage of targets with a negative PG.
Under all other feature sets, for most targets, releasing the synthetic dataset instead of the real data yields a positive PG, meaning that releasing the synthetic dataset instead of the real dataset improves the PG.

\descr{Takeaways.} We find that not even decreasing the attacker's power by only giving him partial information from the target sequence mitigates privacy for the Recomb-generated synthetic data.
This is likely because, using the Recomb model as both generative and  inference model, the adversary's  power is increased since the feature set extracted from the synthetic data will be closer to the feature set for the predicted target.

However, in this case, for RBM, we see an increase in the privacy gained by releasing synthetic data instead of real data.
This implies that, even if the RBM is likely to overfit when few samples are available for training, it does so on the predicted sequence of the target rather than on the full sequence and thus decreases the accuracy of the MIA.

Overall, not even with partial data from the target sequence, we obtain privacy gain values constantly better than random guessing, which, as mentioned before, indicates that synthetic data is not really a reliable privacy defense.

\section{Related Work}\label{sec:related}

In this section, we review related work %
on synthetic data, genomic privacy, and MIAs against machine learning models.

\descr{Synthetic Data Initiatives.}
In recent years, researchers have focused on generating synthetic electronic health records (EHR), aiming to facilitate research in and adoption of machine learning in medicine.
Choi et al.~\cite{choi2017generating} use a combination of
an autoencoder with a GAN model, called medGAN, to generate high-dimensional multi-label discrete data.
ADS-GAN~\cite{YoonDS20} uses a quantifiable definition for ``identifiability'' combined with the discriminator's loss to minimize the probability of patient's re-identification, while CorGAN~\cite{torfi2020corgan} combines convolutional GANs and convolutional autoencoders to capture the correlations between adjacent medical features.
Biswal et al.~\cite{biswal2020eva} use variational autoencoder to synthesize sequences of discrete EHR encounters and encounter features.

Other initiatives focus on synthetic data modeled on primary care data \cite{wang2019generating, tucker2020generating,NhsData20,nhsx_syn_data}.
Researchers have also explored generating synthetic health patient data to detect cancer and other diseases, e.g., RDP-CGAN~\cite{torfi2020differentially} combines convolutional GANs and convolutional autoencoders, both trained with R{\'e}nyi differential privacy \cite{mironov2017renyi}.

Specific to genomes is the work presented in Section~\ref{sec:models}~\cite{SamaniHAEFH15,YelmenDOMMFP19,KilloranLDDF17}, which we have evaluated in terms of their utility--privacy tradeoffs.

\descr{Privacy in Genomics.} Researchers have focused on studying and mitigating privacy risks in genomics.
One of the first attacks on genomic data is the Membership Inference Attack proposed by Homer et al.~\cite{homer2008resolving}, showing that an adversary can infer the presence of an individual's genotype within a complex DNA mixture.
This attack has been improved by Wang et al.~\cite{WangLWTZ09} using correlation statistics of a few hundred SNPs.
Then, Im et al.~\cite{ImGNC12} show that the summary information from genome-wide association studies, such as regression coefficients, can also reveal an individual's participation within the respective study.
Membership inference has also been shown possible in the context of the Beacon network~\cite{ShringarpureB15, RaisaroTJBZCLSBF17,VonThenenAC19}, a federated service that answers queries of the form ``does your data have a specific nucleotide at a specific genomic coordinate?''.

Chen et al.~\cite{Chen2020.08.03.235416} study the effects of differential privacy protection against membership inference attack on machine learning for genomic data. However, their study is focused on privacy leakage via providing access to trained classification models, whereas we study the privacy leakage from sharing synthetic datasets.

\descr{MIAs against Machine Learning Models.}
MIAs have long been studied in the context of machine learning.
Shokri et al.~\cite{ShokriSSS17} present the first %
attack against discriminative models, aiming to identify whether a data record was used in training, using an approach based on shadow models.
Hui et al.~\cite{hui2021practical} also propose an attack against discriminative models, which probes the target model and infers membership directly from the probes instead of shadow models.
Hayes et al.~\cite{HayesMDD19} present the first MIA against generative models like GANs; they use a discriminator %
to output the data with the highest confidence values as the original training data.
Hilprecht et al.~\cite{HilprechtHB19} study MIAs against both GANs and Variational AutoEncoders (VAEs),
while Chen et al. \cite{ChenYZF19} propose a generic MIA model against GANs.

Stadler et al.~\cite{StadlerOT20} recently evaluate MIAs in the context of synthetic data and show that even access to a single synthetic dataset output by the target model can lead to privacy leakage.
We re-use their framework for quantifying the privacy gain when a synthetic dataset is released instead of the real dataset.
However, not only do we do so for a specific context (namely, genomics), but we also measure utility (while they only study privacy).
In fact, there are several distinguishing characteristics between {\em ``generic''} synthetic data and {\em genomic} data, which makes the evaluation significantly different, if not harder; paramount among these is the fact that all features found in genomic sequences are correlated with each other, unlike with generic datasets.
Finally, we introduce and measure a novel attack whereby the attacker only has access to partial genomic information about the target.

\section{Conclusion}
This paper presented an in-depth measurement of state-of-the-art methods to generate synthetic genomic data.
We did so vis-\`a-vis their utility, with respect to several common analytical tasks performed by researchers and the privacy protection they provide compared to releasing real data.

High-quality synthetic data must accurately capture the relations between data points; however, this can enable attackers to infer sensitive information about the training data used to generate the synthetic data.
This was illustrated by the performance of the Recomb model on the HapMap datasets: while it achieves the best utility, it does so at the cost of significantly reducing privacy.

\descr{Discussion.} Overall, we found that there is no single method that outperforms the others for all metrics and all datasets.
While this is perhaps not surprising, we are the first to present a systematic, re-usable methodology to analyze all kinds of methods to generate synthetic genomic data.
For instance, this allows us to be the first to show that, unlike what previously suggested in previous work, models based on a simple GAN architecture (i.e., GAN and Rec-GAN) are not a good fit for genomic data. In fact, these provide low (the lowest) utility across the board.
Our methodology also allowed us to shed light on the influence of the size of the training dataset, especially in the case of generative models.
For example, utility improves wit more samples in the hybrid Rec-RBM for the smaller HapMap datasets, and the RBM for the 1000 Genomes dataset; also, we measured a decrease in the number of targets exposed to membership inference.

We also introduced a new ``MIA with partial information,'' which shed light on the fact that not even decreasing the adversary's power by limiting their knowledge of the target to partial information fully mitigates the privacy loss.
Finally, our modular evaluation framework paves the way for practitioners, scientists, and researchers to easily build on our work and assess the risks of deploying synthetic genomic data in the wild, for a wide range of applications, and serve as a benchmark for techniques proposed in the future.

\descr{Limitations \& Future Work.} %
Our evaluation focuses on existing generative methods for synthetic genomic data; thus, we have not engaged in fine-tuning the (hyper-)parameters of the models evaluated.
Moreover, one might argue that the techniques we evaluate were not designed with privacy in mind, unlike previous work on differentially private generative models for images or clinical data~\cite{acs2018differentially, beaulieu2019privacy, chen2020gs}.
That is, it is not entirely surprising that they yield small privacy gains.
However, to the best of our knowledge, {\em no differentially private generative model has been proposed for genomic data}, which is the focus of our study.

In fact, prior work has shown that, for precision medicine applications, the high dimensionality of the data tends to be a major limitation, resulting in poor utility for differentially private mechanisms~\cite{fredrikson2014privacy, azencott2018machine, jordon2018pate, YaleDRGPB19,xie2018differentially}.
Differentially private techniques for GWAS are also known to yield poor accuracy as the number of features is large, relative to the number of patients in a study~\cite{johnson2013privacy}.
Nonetheless, we plan to experiment with the possible adaptation of differentially private models to genomic data and evaluate them in future work.

Finally, future work should search for, and experiment with, genomics use cases that have more data points, possibly relying on further collaborations with biomedical researchers.
We also intend to extend our privacy evaluation to understand how much the privacy loss stemming from releasing (synthetic) datasets affects the relatives of those included in the training set of the corresponding generative models~\cite{humbert2013addressing,telenti2014genomics}.

\descr{Acknowledgments.} This work was supported by a Google Faculty Award on ``Enabling Progress in Genomic Research via Privacy Preserving Data Sharing.'' 

{\small
\bibliographystyle{abbrv}

}

\begin{figure*}[t]
  \centering
  \subfloat[MAF for CEU]{\includegraphics[width=1.5\columnwidth]{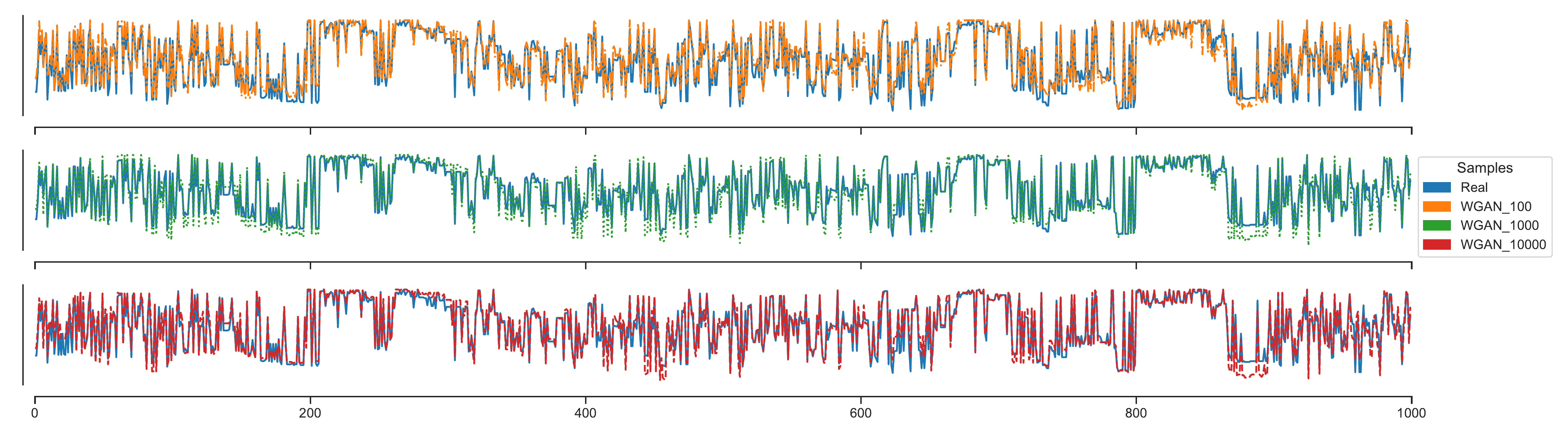}\label{fig:wgans_2}}\\[-0.25ex]
    \subfloat[WGAN discriminator losses for all datasets]{\includegraphics[width=1.8\columnwidth]{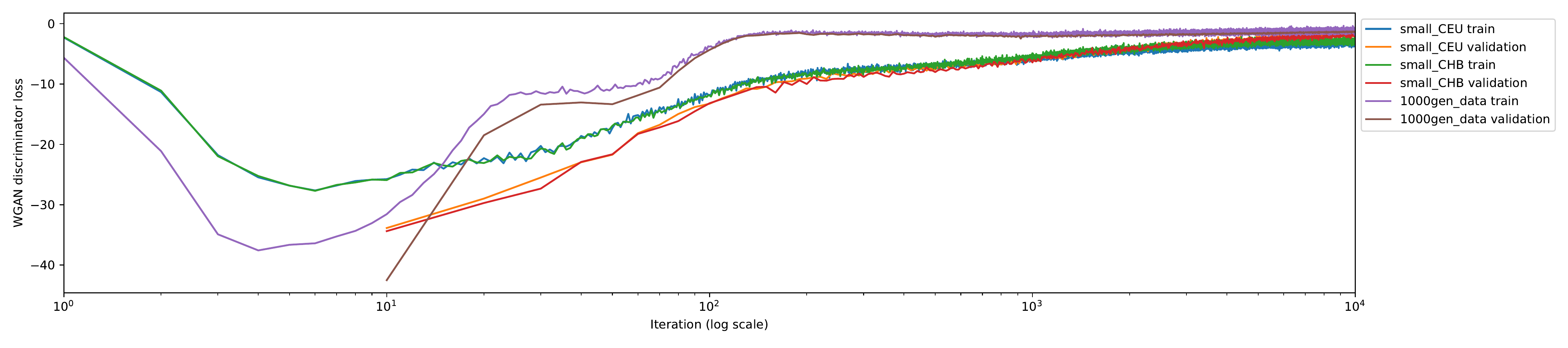}\label{fig:wgans_1}}
  \caption{WGANs experiments for 100/1,000/10,000 iterations.}
  \label{fig:wgans}
  \vspace{-0.15cm}
\end{figure*}
\begin{figure*}[t]
\centering
\subfloat[CEU]{\includegraphics[width=.99\textwidth]{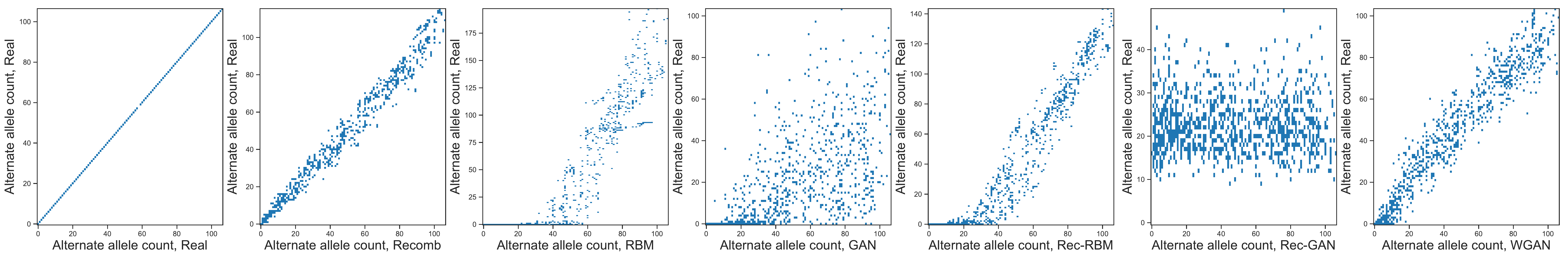}\label{fig:aacceu}}\\[-0.15ex]
\subfloat[CHB]{\includegraphics[width=.99\textwidth]{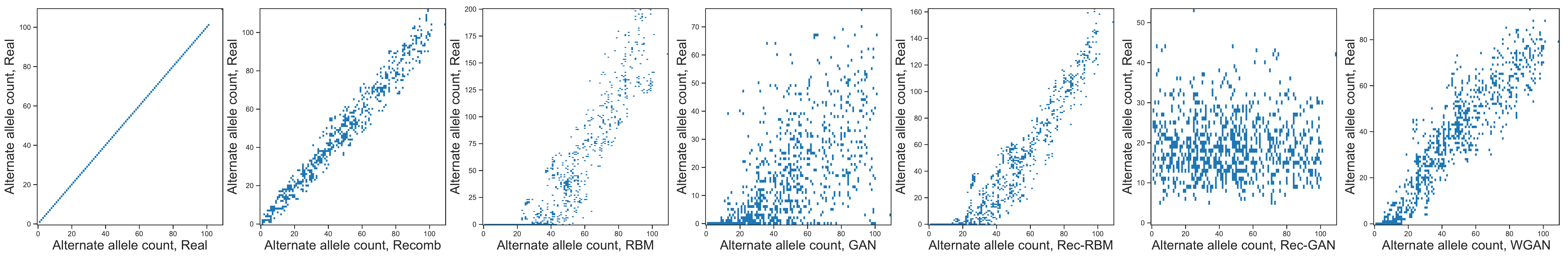}\label{fig:aacchb}}\\[-0.15ex]
\subfloat[1000 Genomes]{\includegraphics[width=.99\textwidth]{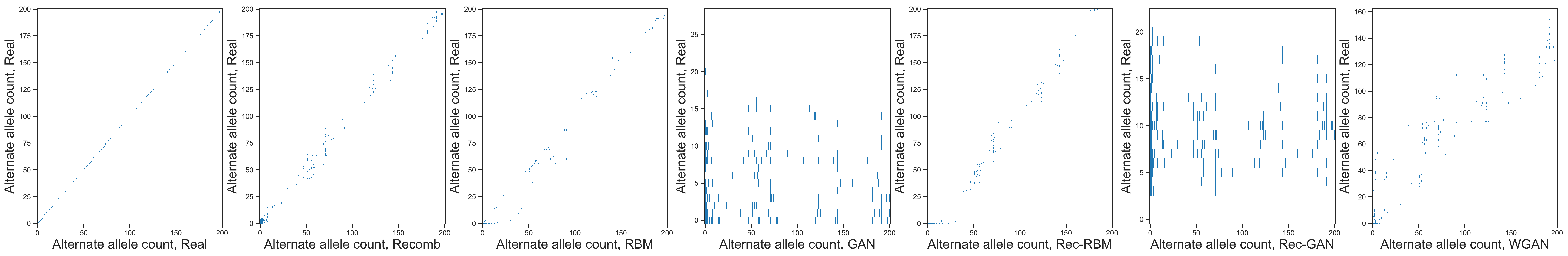}\label{fig:aac1000}}
\caption{Alternate allele correlation for the CEU population, the CHB population, and the 1000 Genomes dataset.}
\label{fig:aac}
\vspace{-0.4cm}
\end{figure*}

\begin{figure*}[t]
\centering
\subfloat[CEU]{\includegraphics[width=0.32\textwidth]{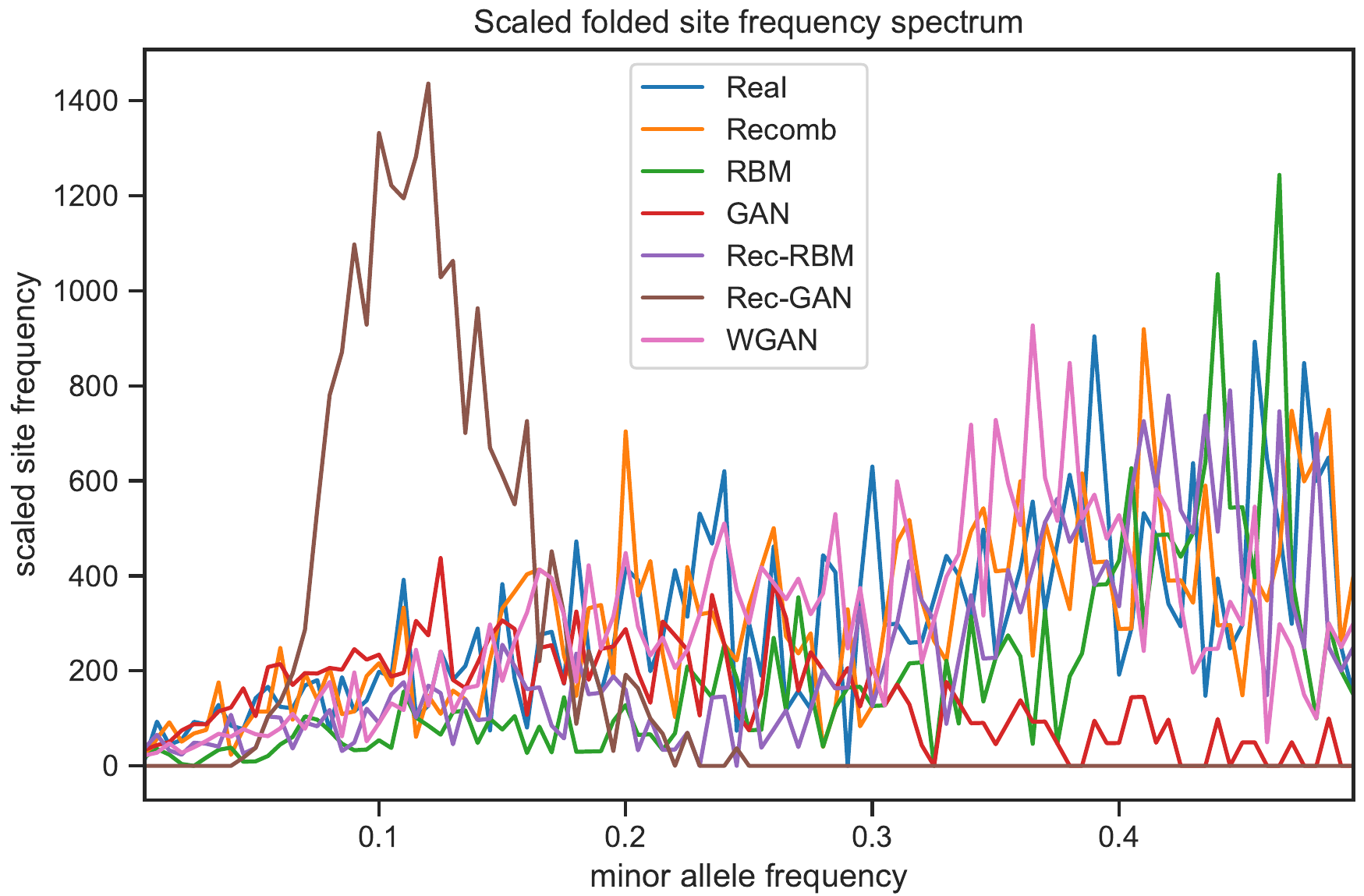}\label{fig:sfsceu}}
\subfloat[CHB]{\includegraphics[width=0.32\textwidth]{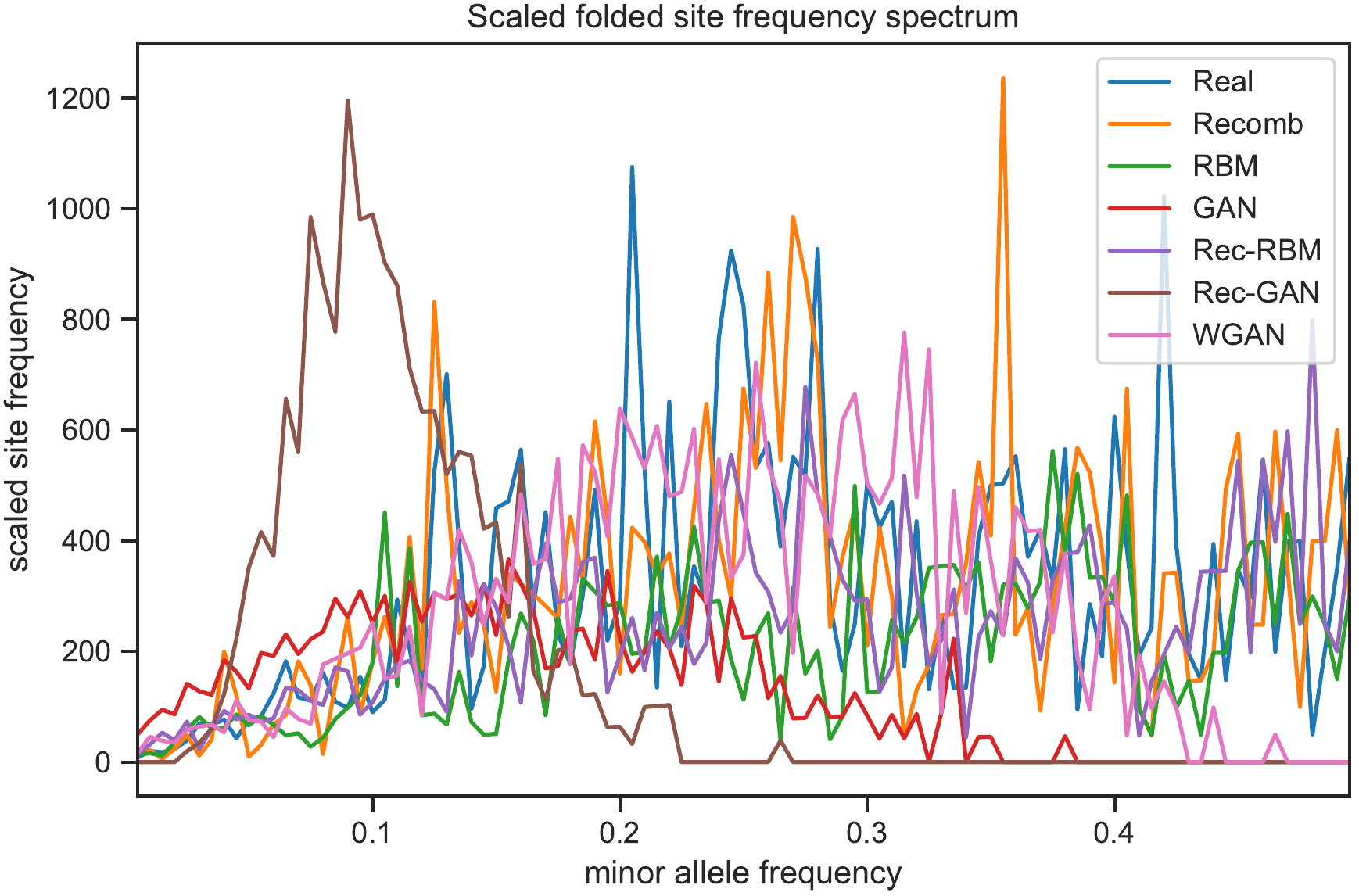}\label{fig:sfschb}}
\subfloat[1000 Genomes]{\includegraphics[width=0.325\textwidth]{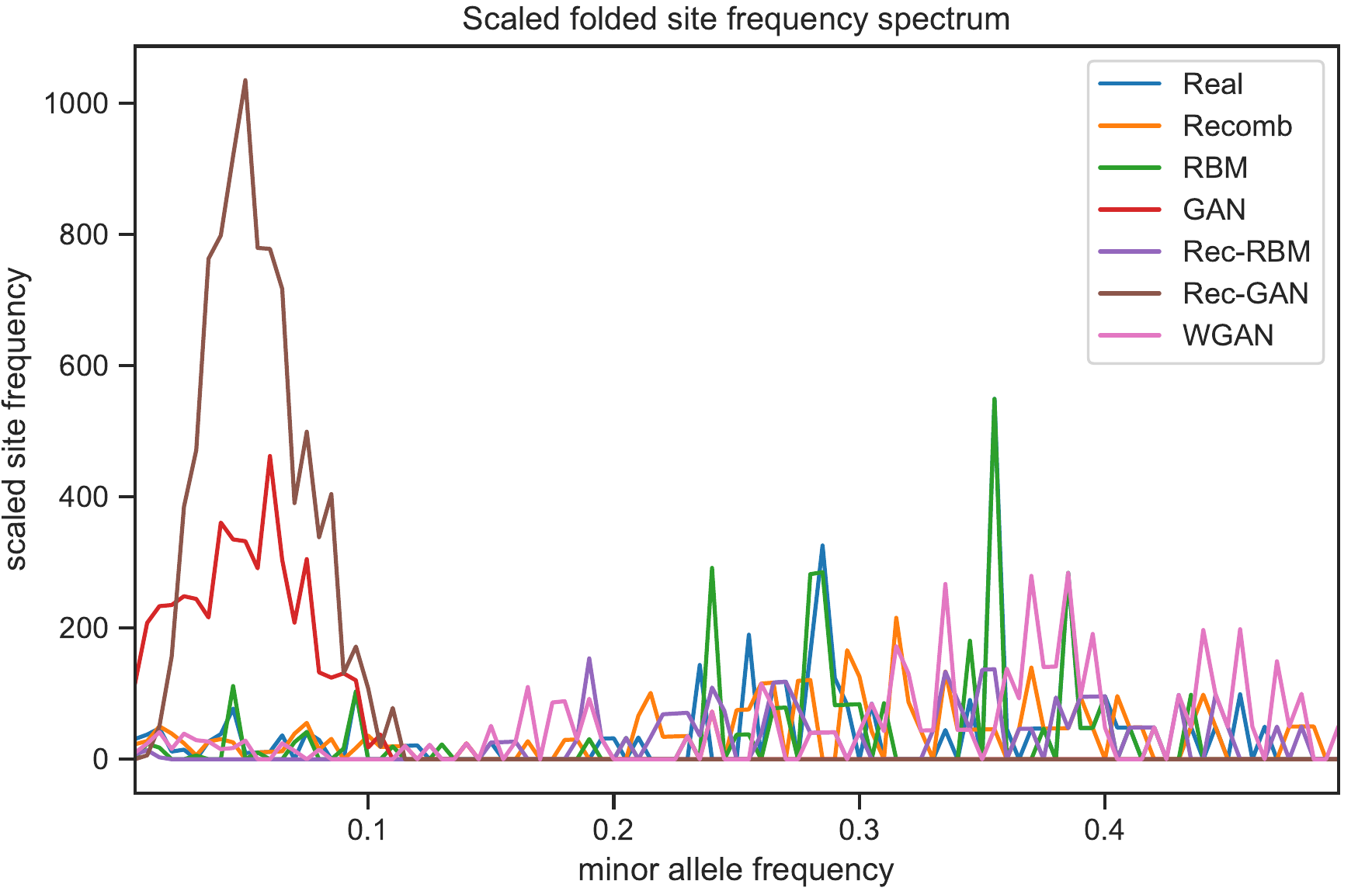}\label{fig:sfs1000}}
\caption{Frequency spectrum analysis for the CEU population, the CHB population, and the 1000 Genomes dataset.}
\label{fig:sfs}
\vspace{-0.2cm}
\end{figure*}

\begin{figure*}[t]
\centering
\subfloat[CEU]{\includegraphics[width=0.99\textwidth]{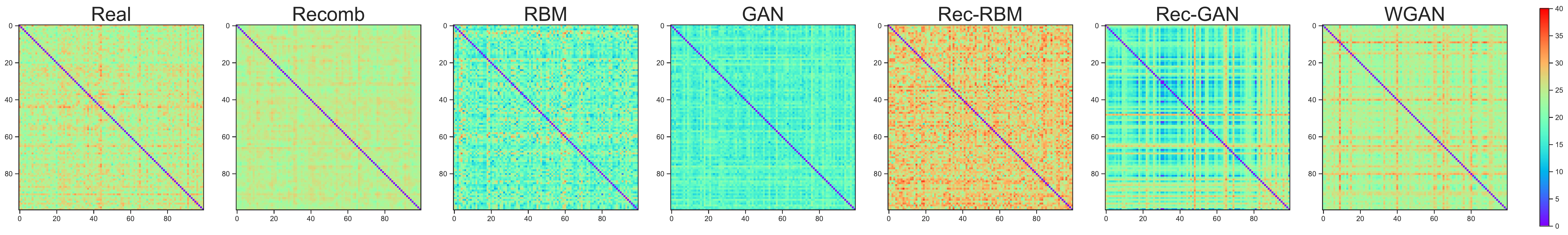}\label{fig:ed-ceu}}\\[-0.15ex]
\subfloat[CHB]{\includegraphics[width=0.99\textwidth]{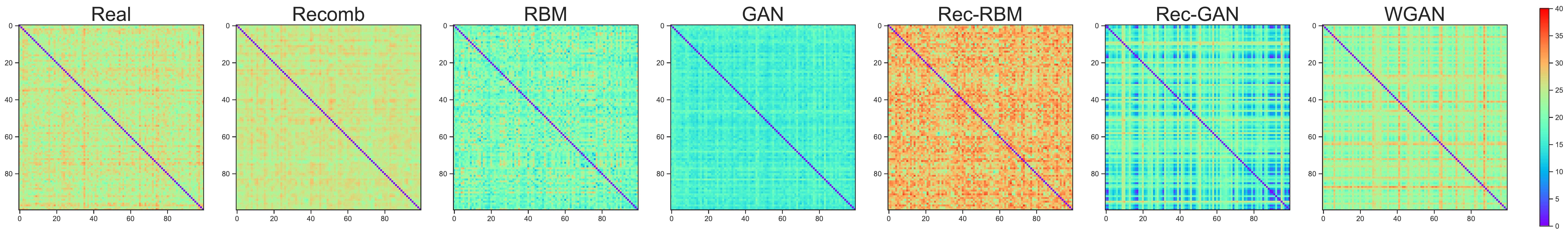}\label{fig:ed-chb}}\\[-0.15ex]
\subfloat[1000 Genomes]{\includegraphics[width=0.99\textwidth]{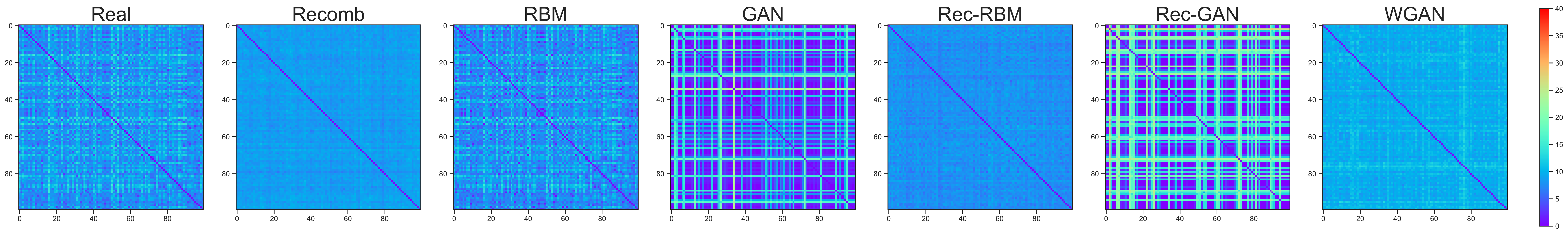}\label{fig:ed-1000}}
\caption{Pairwise Euclidean Genetic Distance (EGD) between individuals.}
\label{fig:ed}
\vspace{-0.4cm}
\end{figure*}

\begin{figure*}[t]
\centering
\subfloat[CEU Population]{\includegraphics[width=1\columnwidth]{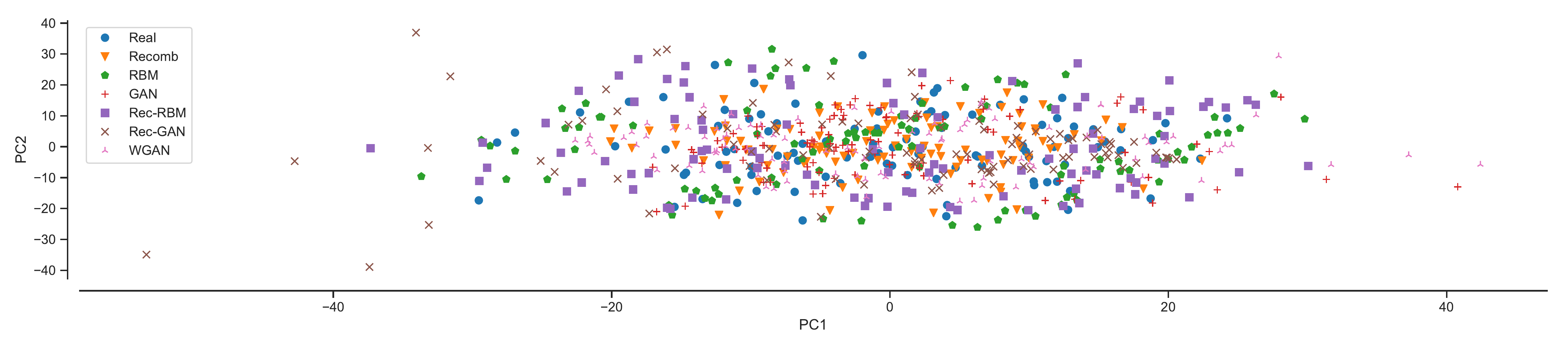}\label{fig:pca-ceu}}
\subfloat[CHB Population]{\includegraphics[width=1\columnwidth]{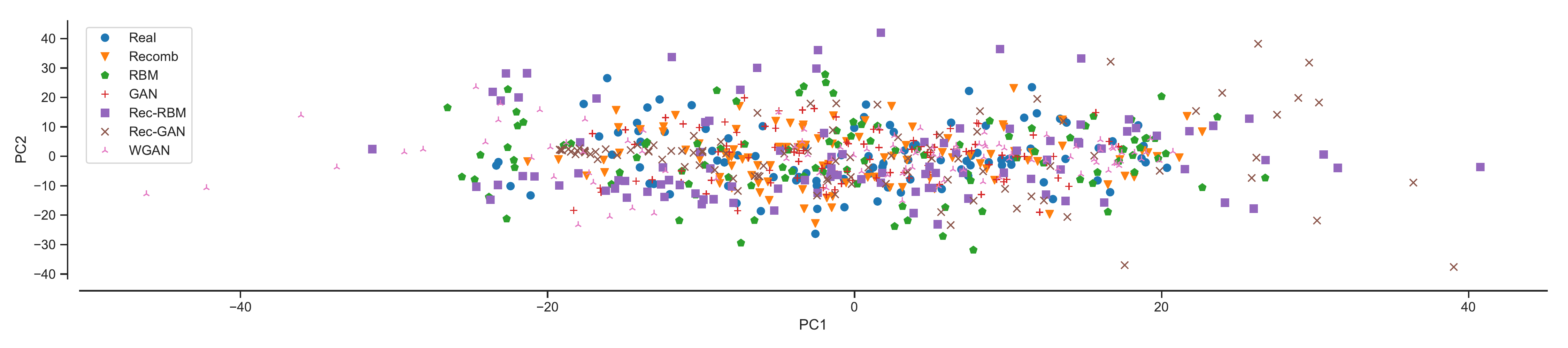}\label{fig:pca-chb}}\\[-0.5ex]
\subfloat[1000 Genomes]{\includegraphics[width=1\columnwidth]{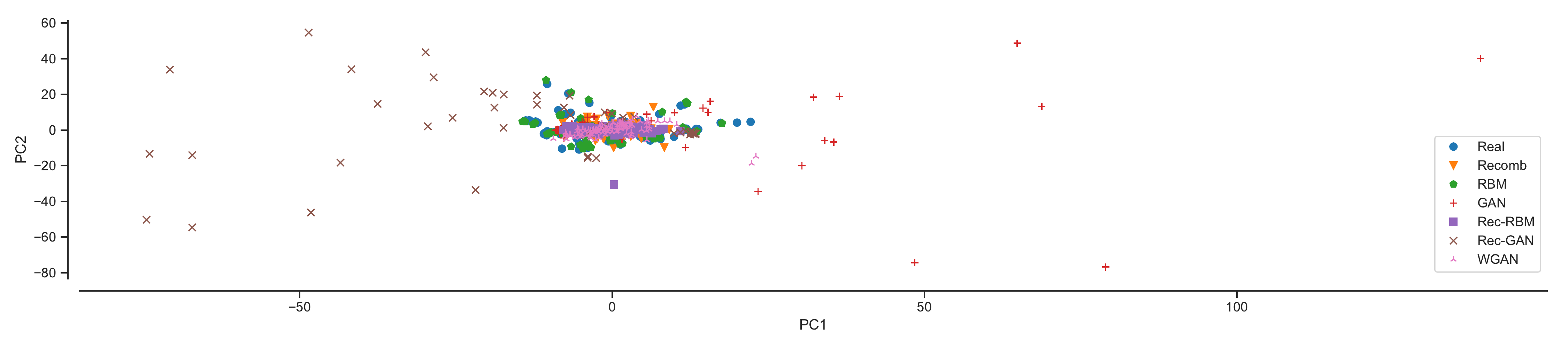}\label{fig:pca-1000}}
\subfloat[1000 Genomes, excluding the GAN and Rec-GAN models]{\includegraphics[width=1\columnwidth]{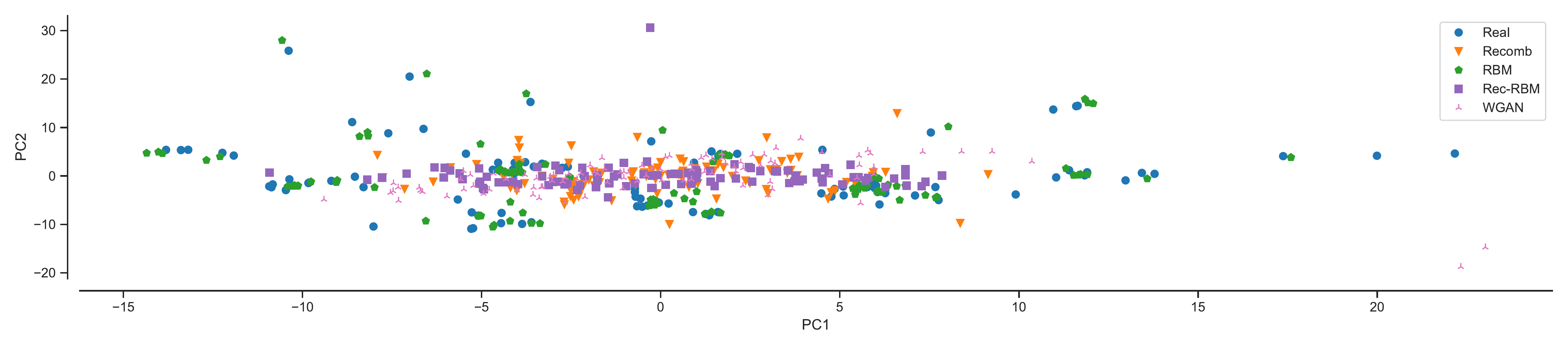}\label{fig:pca-1000-nogan}}\\
\caption{2D Principal Component Analysis (PCA) visualization of the real and synthetic sequences.}
\label{fig:pca}
\end{figure*}

\appendix

\section{WGAN settings}\label{app:wgan}

We experimented with training the WGAN models for various iterations (up to 10,000) and observed that the utility metrics stabilized after around the 100th iteration.
Specifically, it can be seen that for MAF in the CEU dataset in Fig.~\ref{fig:wgans_2} training the WGAN model for more than 100 iterations does not result in better metrics.
The discriminator training and validation losses converge to the same values around 100th iteration for all datasets as can be seen in Fig.~\ref{fig:wgans_1}.
Therefore, in our evaluation we train the WGAN model for 100 iterations, as opposed to the 100,000 proposed in the original implementations.
Overall, we believe this could be due to the characteristics of the datasets as, for example, the sequences in our evaluation all have a length of 1,000.
Killoran et al.~\cite{KilloranLDDF17} experiment with DNA sequences of length around 50.

\section{Additional Figures from Experiments}
In this section, we include some additional experiments complementing our utility evaluation presented in Section~\ref{sec:utility}.
More precisely, Figure~\ref{fig:aac} presents the alternate allele correlation (AAC) between the real and the synthetic datasets, while in Figure~\ref{fig:sfs}, we plot the  scaled folded Site Frequency Spectrum (SFS), which is the distribution of counts of minor alleles in a sample calculated over all segregating sites.
Finally, Figure~\ref{fig:ed} shows the Euclidean Genetic Distance (EGD) between the samples in each dataset.

\section{Principal Component Analysis (PCA)} \label{app:pca}

\begin{figure*}[t]
\centering
\includegraphics[width=0.375\columnwidth]{graphs/plots/legend.png}\\[-2ex]
\subfloat[CEU Population]{\includegraphics[width=0.8\columnwidth]{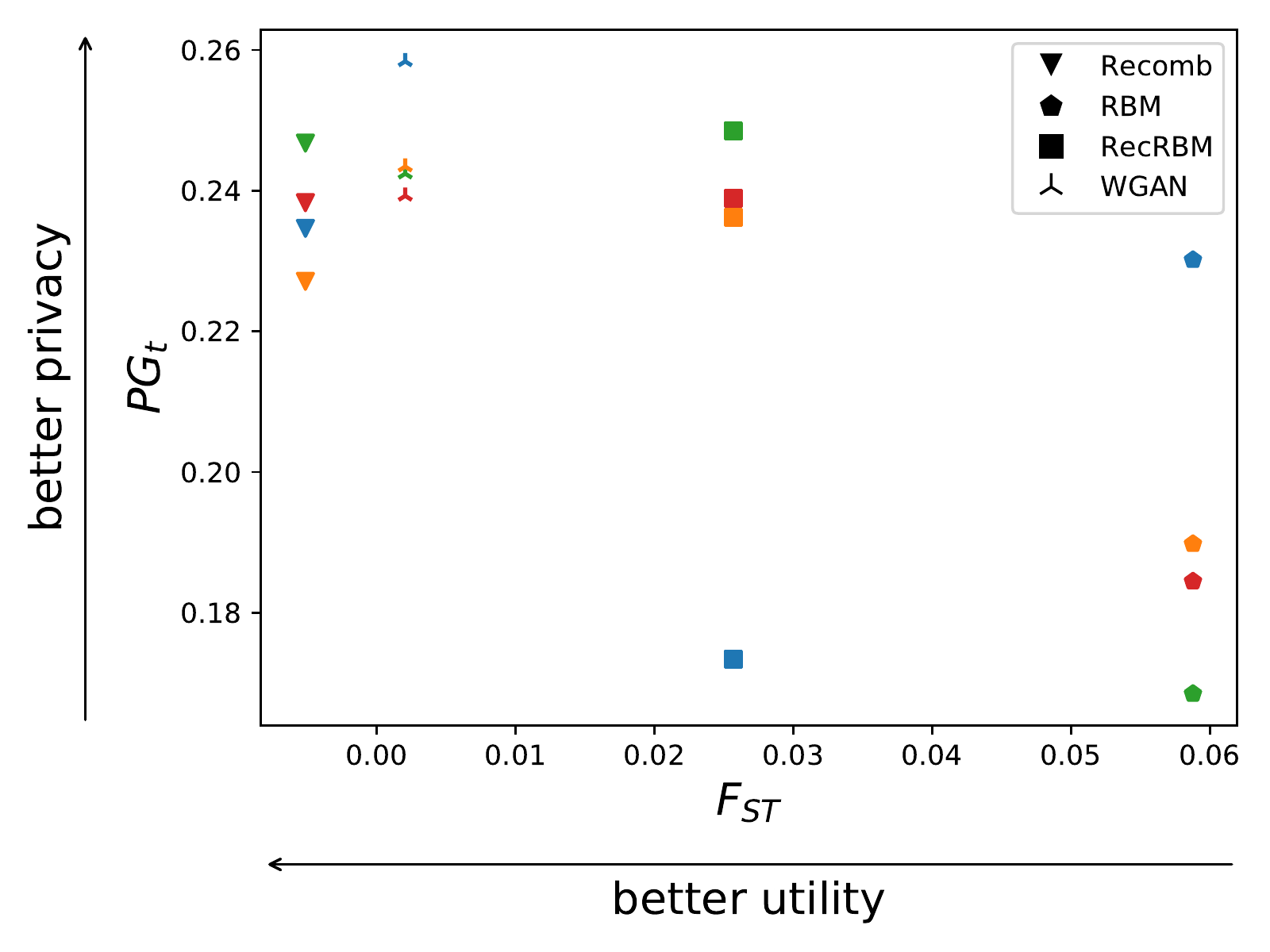}\label{fig:fst_pg_ceu}}
\subfloat[CHB Population]{\includegraphics[width=0.8\columnwidth]{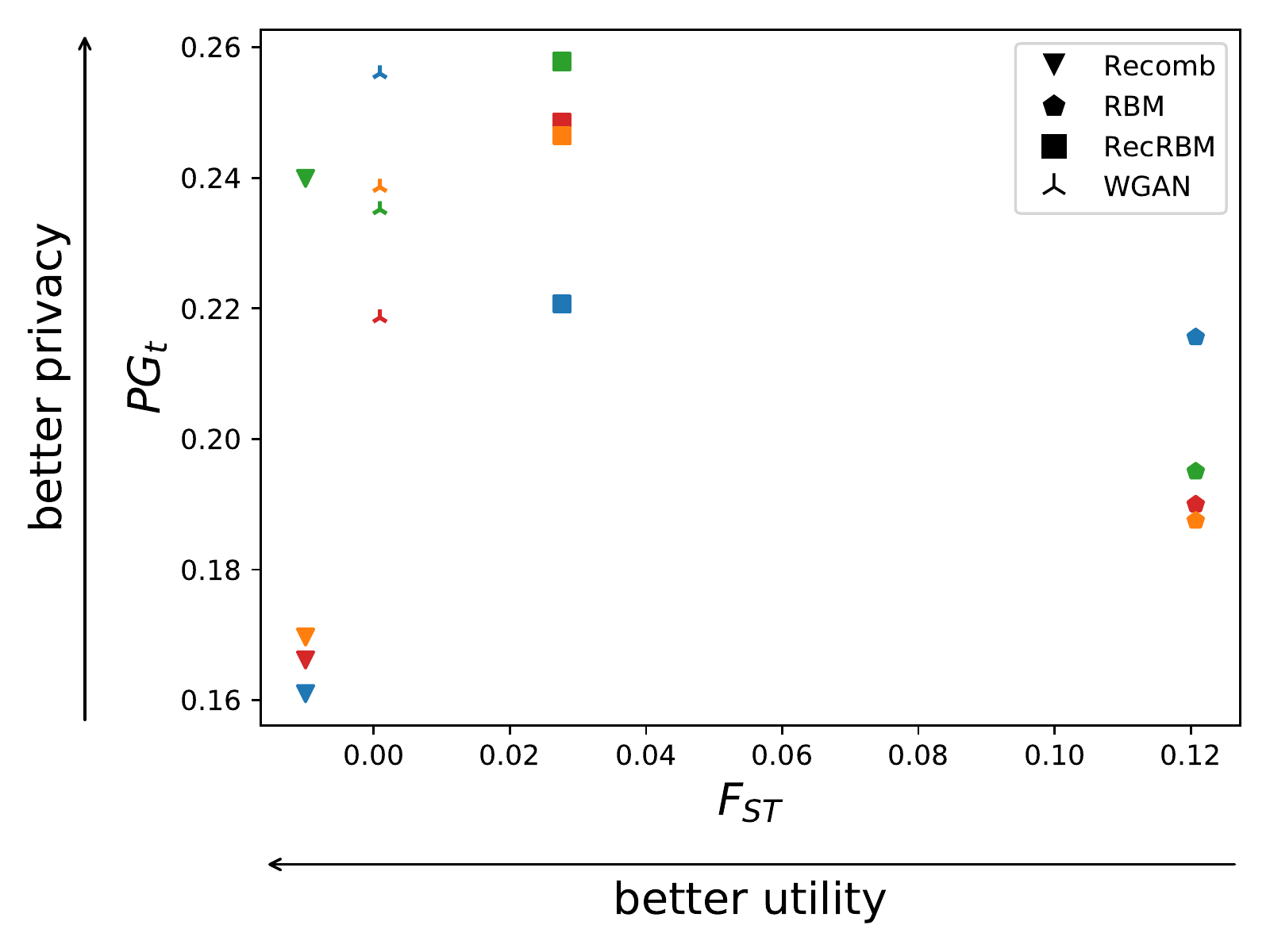} \label{fig:fst_pg_chb}}
\caption{Fixation index values vs Privacy gain for the HapMap populations.}
\label{fig:fst_pg}
\end{figure*}

Here, we study the difference between synthetic and real data vis-\`a-vis a principal component analysis on the corresponding samples.
We extract the first two principal components and project the real and synthetic datasets on these two components to show how the synthetic samples are distributed compared to the real data.

Fig.~\ref{fig:pca} presents this 2D visualization.
Recomb has a close distribution to the real data for both HapMap populations, which, according to \cite{SamaniHAEFH15}, is because the genetic recombination model considers all the correlations between SNPs and builds a higher-order model. %
For the 1000 Genomes, as for the other utility metrics studied in Section~\ref{sec:utility}, the GAN and Rec-GAN models perform quite poorly, generating samples with a different distribution than the real samples.
Therefore,  in Figure~\ref{fig:pca-1000-nogan}, we exclude them in order to take a closer look at the non-GAN-based models.
Here, we can better observe that the RBM-generated samples have the closest distribution to the real data.
In contrast to the HapMap populations, the samples from Recomb are all centered around 0 and fail to simulate the distribution given by the real data, and similar results are in the case of samples generated by Rec-RBM.

\section{Privacy Gain vs. $F_{ST}$}\label{app:tradeoff}

To provide a quick visualization of the {\em trade-offs} between privacy and utility, we also plot the $F_{ST}$ (see Section~\ref{subsec:pop_stat}) against the PG in Figure~\ref{fig:fst_pg} for the two smaller HapMap Populations.
For this set of experiments, we randomly sample 10 targets from each dataset.
We train a Logistic Regression attack classifier using 5 shadow models, using 100 synthetic datasets for each of them.
We then compute the PG and $F_{ST}$ on 100 synthetic datasets, with a split of 50 sets generated from a training set including the target and 50 sets generated without.
We report the PG and $F_{ST}$ as the average across all targets and synthetic datasets.

Recall that the lower the $F_{ST}$ value the closer the synthetic samples are to the real data, and PG should be above the random guess baseline of 0.25 for the synthetic data to offer better privacy protection than releasing the real.
Put simply, the ideal ``place'' in the plots in Figure~\ref{fig:fst_pg} is the top left.

For the CEU population (Figure~\ref{fig:fst_pg_ceu}), the synthetic Recomb samples have the best overall utility; however, the average PG is below 0.25.
In line with our utility evaluation, we find that the hybrid model Rec-RBM model generates samples with a $F_{ST}$ value closer to the real data than the RBM model.
Interestingly, while for Histogram, Correlation, and Ensemble feature sets, the PG for the Rec-RBM is close to 0.25, there is a significant privacy loss for Naive.
In contrast, for the samples generated by the RBM, the Naive yields the highest PG across all feature sets.
The samples generated by the WGAN also have a $F_{ST}$ value close to 0; however, for the Histogram, Correlation and Ensemble, the PG values are below 0.25.

For CHB (Figure~\ref{fig:fst_pg_chb}), the utility results are similar to the CEU population.
However, all targets apart from the ones generated by RecRBM have a lower PG than the CEU population.
This reiterates the fact that one cannot reliably use synthetic data as a good privacy mechanism, as the value of PG is unpredictable and can fluctuate based on target and training set combinations chosen.

\end{document}